\documentclass[a4paper,fleqn]{cas-dc}

\usepackage[numbers]{natbib}
\usepackage{color}

\def\tsc#1{\csdef{#1}{\textsc{\lowercase{#1}}\xspace}}
\tsc{WGM}
\tsc{QE}
\tsc{EP}
\tsc{PMS}
\tsc{BEC}
\tsc{DE}
\begin{document}
\let\WriteBookmarks\relax
\def\floatpagepagefraction{1}
\def\textpagefraction{.001}
\shorttitle{Neural Architecture Search for Compressed Sensing Magnetic Resonance Image Reconstruction}
\shortauthors{J. Yan et~al.}

\title [mode = title]{Neural Architecture Search for Compressed Sensing Magnetic Resonance Image Reconstruction}           

\author[1,3]{Jiangpeng Yan}[ orcid=0000-0002-0767-1726]
\credit{Conceptualization of this study, Methodology, Software, Writing - Original draft preparation}
\address[1]{Department of Automation, Tsinghua University, Beijing 100091, China}

\author[2]{Shou Chen}
\credit{Data curation, Validation,  Writing - Review and editing}
\address[2]{Center for Biomedical Imaging Research, Department of Biomedical Engineering, School of Medicine, Tsinghua University, Beijing 100091, China}

\author[3]{Yongbing Zhang}
\credit{Methodology, Writing - Review and editing}

\author[3]{Xiu Li}[orcid=0000-0003-0403-1923]
\cormark[1]
\ead{li.xiu@sz.tsinghua.edu.cn}
\credit{Project administration, Supervision, Writing - Review  and editing, Funding acquisition}
\address[3]{Tsinghua Shenzhen International Graduate School, Tsinghua University, Shenzhen 518055, China}

\cortext[cor1]{Corresponding author}

\begin{abstract}
Recent works have demonstrated that deep learning (DL) based compressed sensing (CS) implementation can accelerate Magnetic Resonance (MR) Imaging by reconstructing MR images from sub-sampled k-space data. However, network architectures adopted in previous methods are all designed by handcraft. Neural Architecture Search (NAS) algorithms can automatically build neural network architectures which have outperformed human designed ones in several vision tasks. Inspired by this, here we proposed a novel and efficient network for the MR image reconstruction problem via NAS  instead of manual attempts. Particularly, a specific cell structure, which was integrated into the model-driven MR reconstruction pipeline, was automatically searched from a flexible pre-defined operation search space in a differentiable manner. Experimental results show that our searched network can produce better reconstruction results compared to previous state-of-the-art methods in terms of PSNR and SSIM with $4\sim6$ times fewer computation resources. Extensive experiments were conducted to analyze how hyper-parameters affect reconstruction performance and the searched structures. The generalizability of the searched architecture was also evaluated on different organ MR datasets. Our proposed method can reach a better trade-off between computation cost and reconstruction performance for MR reconstruction problem with good generalizability and offer insights to design neural networks for other medical image applications. The evaluation code will be available at \url{https://github.com/yjump/NAS-for-CSMRI}.
\end{abstract}

\begin{highlights}
\item Neural Architecture Search is firstly integrated into the compressed sensing MR image reconstruction task instead of manual designed networks.
\item Better reconstruction results are achieved with $4\sim6$ times fewer computation resources than baseline methods.
\item The extensive analysis of how hyper-parameters affect reconstruction results and the searched structures, may offer insights to design neural networks for other medical analysis applications.
\item The searched architectures can be directly generalized to different organ MR reconstruction tasks by re-training and will be more task-specific via re-searching.
\item The evaluation code will be available at \url{https://github.com/yjump/NAS-for-CSMRI}.
\end{highlights}


\maketitle

\section{Introduction}
\label{sec:introduction}
Magnetic Resonance Imaging (MRI) can noninvasively provide various contrast images for assessing anatomical structures and physiological functions. However, the raw data of MRI are acquired in k-space (i.e. Fourier space), and the speed of scanning is limited by physiological constraints \cite{liang2000principles}. The relative slow imaging speed limits the use of MRI in applications that are motion-sensitive or time-starved. Thus, MRI acceleration has been an active research area since it was proposed in the 1970s \cite{tsao2012mri}.

\begin{figure*}
\centerline{\includegraphics[width=14cm]{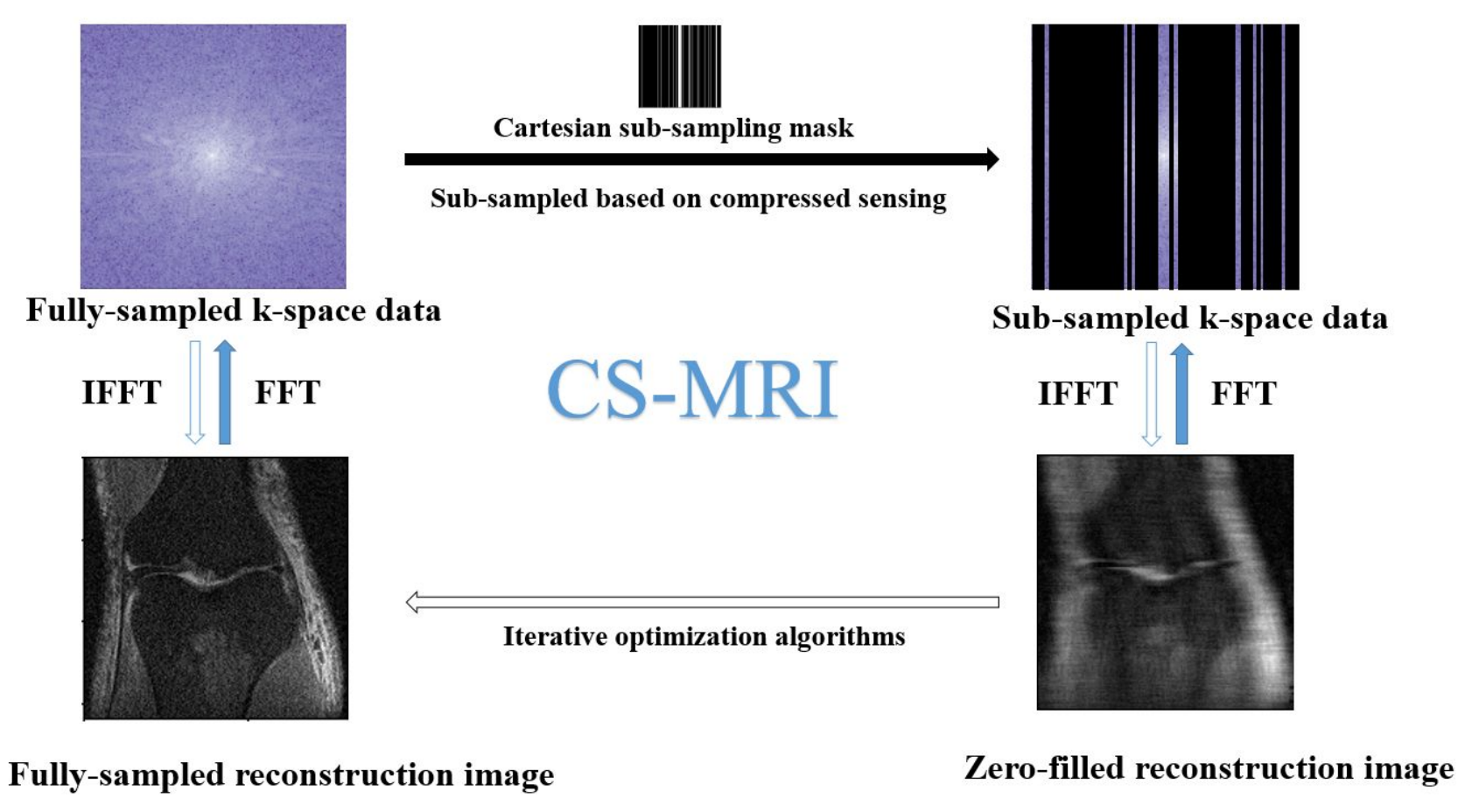}}
\caption{ An illustration of the traditional CS-MRI acceleration framework. Firstly, the k-space signals acquisition process is accelerated by sub-sampling. Then, zero-filled reconstruction images are achieved by performing Inverse Fast Fourier Transform (IFFT) on sub-sampled k-space data. Finally, high quality MR images can be reconstructed by alleviating aliasing patterns in the zero-filled reconstruction images using iterative optimization algorithms. For example, if an 8-fold Cartesian sub-sampled mask is applied in k-space data, the signals acquisition process can be accelerated 8 times theoretically.}
\label{fig1}
\end{figure*}

Among various MRI acceleration methods, Compressed Sensing Magnetic Resonance Imaging (CS-MRI) gains a lot of attention because it can significantly accelerate MRI without any additional hardware \cite{lustig2008compressed}. The key principle of compressed sensing \cite{Donoho2006Compressed} is that we can reconstruct high-quality images from sub-Nyquist sampling signals when the following two assumptions are satisfied: first, the image has a sparse representation in a specific transform domain; second, the sampling and the sparsity domain are incoherent. Based on this theory, we can reconstruct MR images from randomly sub-sampled k-space data by iterative optimization algorithms to suppress the aliasing artifacts, i.e. incoherent noise, caused by sub-Nyquist sampling. Thus, the major problem of CS-MRI transfers to: 1) designing a better sparse representation \cite{liu2017sparse} for de-aliasing; 2) the efficient implementation for clinical translation. The whole framework of traditional CS-MRI is shown in Fig.\ref{fig1}.

Nowadays, Deep Learning (DL) \cite{Lecun2015Deep} has achieved a dominant position in many computer vision applications including object detection \cite{Girshick2014Rich}, semantic segmentation \cite{Long2015FullyCN},
image de-noising \cite{Xie2012Image}, image super-resolution\cite{Dong2014ImageSU}, etc. Researchers have also validated the feasibility of deep neural networks based MRI reconstruction. As far as we know, the deep convolutional neural network (CNN) was firstly introduced to CS-MRI by Wang et al. \cite{Wang2016Accelerating}, where a three-layer CNN was trained with L2 loss between paired zero-filled and fully-sampled reconstruction MR images. Since then, significant progresses have been made by researchers to produce better MR image reconstruction results including data-driven methods \cite{zhu2018automap, mardani2018gan} and model-driven methods \cite{Schlemper2017ADC, aggarwal2018modl}. A detailed review will be introduced in the following section. Researchers have successfully developed various DL based frameworks for CS-MRI, but limited attention is paid to how the network architectures can affect the reconstruction process. In previous works, all the networks were designed by handcraft, thus the performances of these networks highly depend on researchers’ expertise and labor with the following two concerns. On one hand, only several common convolutional operations (e.g. $3 \times 3$ convolution) are tried in current works and other operations (e.g. dilated convolution) with their possible combinations are not sufficiently explored. On the other hand, it is hard to balance the performance and computation cost of CNNs by limited manual attempts. 

Recently, more and more researchers focus on developing
algorithmic solutions to automate the manual process
of architecture design. Architectures automatically found
by Neural Architecture Search (NAS) algorithms have achieved better performance with fewer computation resources in various vision tasks such as image classification \cite{zoph2016neural, cai2018efficient, liu2018darts} and semantic segmentation
\cite{liu2019auto}. Inspired by this, here we proposed a novel and efficient MR image reconstruction network by NAS. Our main contributions can be summarized as follows:

\begin{itemize}
    \item We present a novel network designed for MR image reconstruction via NAS instead of manual attempts. To the best of our knowledge, we are the first to introduce NAS to solve CS-MRI problem.
    \item Experimental results on a knee MR dataset demonstrate that the searched network can achieve better performance with $4\sim6$ times fewer computation resources than manually designed ones. The combination of NAS and CS-MRI is effective.
    \item How hyper-parameters affect reconstruction results and the searched structures was explored, which may offer insights to design neural networks for other medical image applications.
    \item Extensive experiments on a brain MR dataset prove that the searched network can be directly generalized to different organs via re-training. And a more task-specific architecture can be identified by re-searching. 
    \item  The evaluation code will be available at \url{https://github.com/yjump/NAS-for-CSMRI}.
\end{itemize}

The organization of this paper is as follows. In Section II, recent developments of DL based CS-MRI frameworks and NAS algorithms are reviewed. In Section III, details of how to search the network architecture are elaborated. In Section IV, experimental results show that our model is effective and efficient from both the quantitative and qualitative perspective. Extensive experiments of hyper-parameters and model generalizability are also presented. In Section V, we discuss and draw the conclusion of this study.

\section{Related work}

\subsection{Recent Developments in DL based CS-MRI frameworks}
\label{sec:csmri}

Current DL based CS-MRI frameworks can be roughly divided into two categories: data-driven and model-driven methods. 

For data-driven methods, inspired by the initial work of Wang et al. \cite{Wang2016Accelerating},  researchers proposed different frameworks to learn the relationship between sub- and fully-sampling data. To build the mapping from sub-sampled k-space data to fully-sampled image data, Zhu et al. \cite{zhu2018automap} proposed AutoMap model with fully-connected and convolutional layers. To estimate the missing k-space data, RAKI \cite{akccakaya2019LAKI} and LORAKI \cite{kim2019loraki} focused on using CNNs to implement interpolation reconstruction in the k-space domain. To perform reconstruction in the image domain, Lee et al.\cite{lee2018deep} reconstructed the magnitude and phase of MR images by two separated networks. Different generative adversarial networks (GAN) \cite{Goodfellow2014Generative} were explored in \cite{mardani2018gan,Yang2017DAGAN,Quan2018CyclicLoss} to reconstruct MR images with a discriminator-based loss for recovering more detailed textures. In these data-driven methods, deep networks can be regarded as the ``black box" trained in an end-to-end fashion from the input domain to the output domain directly.

For model-driven methods, researchers used deep networks to learn image priors for MR image reconstruction and then integrated these networks into traditional algorithms to unroll the iterative optimization process. Sun et. al. \cite{sun2016ADMM} firstly used convolutional layers to unroll the Alternating Direction Method of Multipliers (ADMM) optimization to solve the single-coil MR image reconstruction. A variational network was proposed by \cite{hammernik2018vn} to deal with the multi-coil problem. The data consistency module proposed in DCCNN \cite{Schlemper2017ADC} that performs iterative reconstruction in a cascading way has a great impact on the following works \cite{Huang2018MRIRV, Sun2018CompressedSM, Zeng2019AVD} and was generalized for common inverse problems and multi-channel MR data as the Model-based Deep Learning (MoDL) \cite{aggarwal2018modl}.

Although there exist various reconstruction frameworks now, the network architectures used in both data-driven and model-driven methods are very similar. U-net \cite{Ronneberger2015U} and its variants with residual learning \cite{lee2018deep}, cascading n-fold architecture \cite{Quan2018CyclicLoss} and channel-wise attention \cite{Huang2018MRIRV} were explored in independent works. U-net is famous for its success in medical image semantic segmentation, but it is not designed specifically for MR image reconstruction. In \cite{Wang2016Accelerating, Schlemper2017ADC, aggarwal2018modl}, plain fully convolutional networks were adopted. Following DCCNN \cite{Schlemper2017ADC}, RDN \cite{Sun2018CompressedSM} introduced dilated convolution \cite{Yu2015MultiScaleCA} and recursive learning \cite{Kim2015DeeplyRecursiveCN} to produce higher quality reconstructions with fewer parameters. Huang et al.\cite{Huang2018MRIRV} and Zeng et al.\cite{Zeng2019AVD} also focused on how to design fine and novel structures instead of plain CNN with data consistency modules to improve final results. These architectures are all manually designed and have limited performance due to the concerns mentioned above. In contrast to these works, we tried to find a better architecture with the help of NAS.

\subsection{Recent Developments in NAS Algorithms}

The network architecture plays an important role in the study of DL and there are many famous architectures, e.g. AlexNet \cite{krizhevsky2010convolutional}, InceptionNet \cite{szegedy2017inception}, ResNet \cite{He2016Deep}, etc. NAS aims to develop algorithms for automatical neural architectures design. At a high
level, current methods usually fall into three categories: evolutionary algorithm (EA) \cite{angeline1994evolutionary}, reinforcement learning (RL) \cite{sutton1998introduction}, and differentiable search \cite{liu2018darts}. 

In EA based NAS methods\cite{real2017large, real2019regularized}, the best architecture was obtained by progressively mutating a population of candidate architectures. Reinforcement learning (RL) technique is an alternative to EA in \cite{zoph2016neural, cai2018efficient, liu2018progressive}  by training a recurrent neural network \cite{mikolov2010recurrent} meta-controller to generate final architectures from a predefined sequences encoding search space.
The major limitation of these EA and RL based methods is that they tend to require a large number of computation resources.

Our work is most closely related to the final differentiable search methods. 
Based on the continuous relaxation of the architecture representation \cite{liu2018darts}, the architecture of inner cells can be selected via back propagation \cite{LeCun1989HandwrittenDR} automatically. Recent applications of differentiable search all focused on the classification and segmentation task of natural images, typically DARTS \cite{liu2018darts} and Auto-Deeplab\cite{liu2019auto}. These automatically searched network architectures have outperformed previous handcrafted ones in these tasks.

\section{Methodology}

To clarify our method, we firstly introduce DL based MR image reconstruction framework, then we present our neural architecture search strategy. Because multi-channel MRI data need a great number of computation resources, we perform all the formulations and experiments in the single-coil MR image reconstruction scene. 

\subsection{DL based MR Image Reconstruction Framework}
\label{sec:form}
We followed DCCNN \cite{Schlemper2017ADC} and MoDL\cite{aggarwal2018modl} to unroll the
alternating minimization algorithm with cascading CNN-derived
constraint for the CS-MRI problem.  

The aim of CS-MRI is to reconstruct the fully-sampled, i.e. aliasing free, MR image $x \in \mathbb{C}^{N}$ from the sub-sample k-space measurement $s \in \mathbb{C}^{M} ( M < N ) $, such that:

\begin{equation}
\label{equ:2}
s = U x,
\end{equation}
where $U \in \mathbb{C}^{M \times N}$ is the sub-sampling encoding matrix (e.g. Fourier encoding). Then $x$ can be obtained by solving the following unconstrained optimization problem:

\begin{equation}
\label{equ:3}
\min_x \lVert s - U x\rVert_2^2 + \mathcal{R}(x),
\end{equation}
where $\mathcal{R}(x)$ is the regularization term in the image domain, and $\lVert s - U x\rVert_2^2$ can be regarded as the data consistence term between the image domain and k-space domain. For the traditional CS-MRI framework, L1 or L2 norm of $x$ is often used as the regularization term. For DL based CS-MRI,  the deep CNN is integrated into this formulation by:

\begin{equation}
\label{equ:4}
\min_x  \lVert s - U x\rVert_2^{2} + \lambda \lVert x - \mathcal{D} \left( x , \omega )\right \rVert^2,
\end{equation}
where $\mathcal{D}$ represents the deep CNN network with learn-able weights $\omega$. This problem can be solved with the alternating minimization steps by:

\begin{equation}
\label{equ:k}
y^{(n)} = \mathcal{D}( x^{(n)} , \omega ).
\end{equation}

\begin{equation}
\label{equ:m}
x^{(n+1)} = \arg\min_x  \lVert s - U x\rVert_2^{2} + \lambda \lVert x - y^{(n)}\rVert^2.
\end{equation}

The sub-problem of Eq.\ref{equ:k} can be regarded as a de-aliasing problem in the image domain. Given paired zero-filled MR reconstruction image as $x$ and fully-sampled MR image $y$, the CNN $\mathcal{D}$ with its weights $\omega$ can be obtained by minimizing the objective function:

\begin{equation}
\label{equ:5}
\mathop{{\min}}\limits_{{\omega}} \mathcal{L} (\mathcal{D}(x, \omega)-y),
\end{equation}
where $\mathcal{L}$ is the loss function, e.g. L1 loss or L2 loss.

The sub-problem of Eq.\ref{equ:m} is related to the data consistency problem between k-space and image domain. As for single channel MR image acquisition scenario, i.e. $U = MF$ where $F \in \mathbb{C}^{N \times N}$ applies two-dimensional Fast Fourier Transform (FFT) and $M \in \mathbb{C}^{M \times N}$ is the sub-sampling mask selecting lines in k-space, Eq.\ref{equ:m} has a close-formed solution:

\begin{equation}
\label{equ:cls}
x = (1 + \lambda  U^\mathrm{T}U)^{-1}(\mathcal{D}(x, \omega) +\lambda  U^\mathrm{T}s).
\end{equation}
This solution was firstly introduced in DCCNN \cite{Schlemper2017ADC} as the data consistency process. MoDL\cite{aggarwal2018modl} generalized Eq.\ref{equ:m} to multi-channel acquisition cases. For example, to deal with multi-coil MR image, we have $U = MFS$ where the coil sensitivity map $S$ needs to be taken into consideration so the conjugate gradient (CG) algorithm is used to solve this more complex problem because $1 + \lambda  U^\mathrm{T}U$ is not analytically invertible. Then, the data consistency process can be integrated as a layer with deep CNNs without trainable parameters. In other 
words, we fuse the accurate partial k-space data into deep CNNs to correct biases that accrue during the inference periods by Eq.\ref{equ:cls}. 

\begin{figure*}
\centerline{\includegraphics[width=14cm]{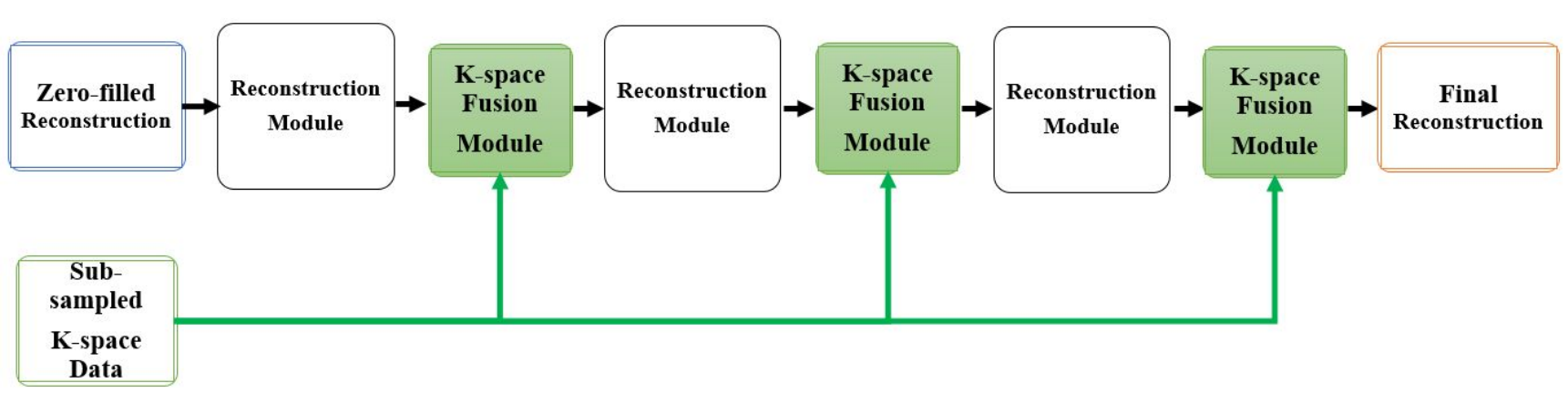}}
\caption{ The backbone framework of MR image reconstruction. The K-space Fusion Module helps to correct biases that accrue during the inference of the CNN. This strategy is widely used in different works \cite{Schlemper2017ADC, aggarwal2018modl, Huang2018MRIRV,  Sun2018CompressedSM, Zeng2019AVD}, with various reconstruction modules. In MoDL\cite{aggarwal2018modl}, all the reconstruction modules share the same weights to reduce the number of learn-able parameters.}
\label{fig2}
\end{figure*}

Note the deep CNNs by reconstruction modules and the data consistency process by k-space data fusion modules, we can unroll Eq.\ref{equ:m} and Eq.\ref{equ:k} by cascading these modules. According to the limited computation resource we have, we iterate these modules three times to form the final backbone shown in Fig.\ref{fig2}.

\begin{figure}[ht]
\centerline{\includegraphics[width=9cm]{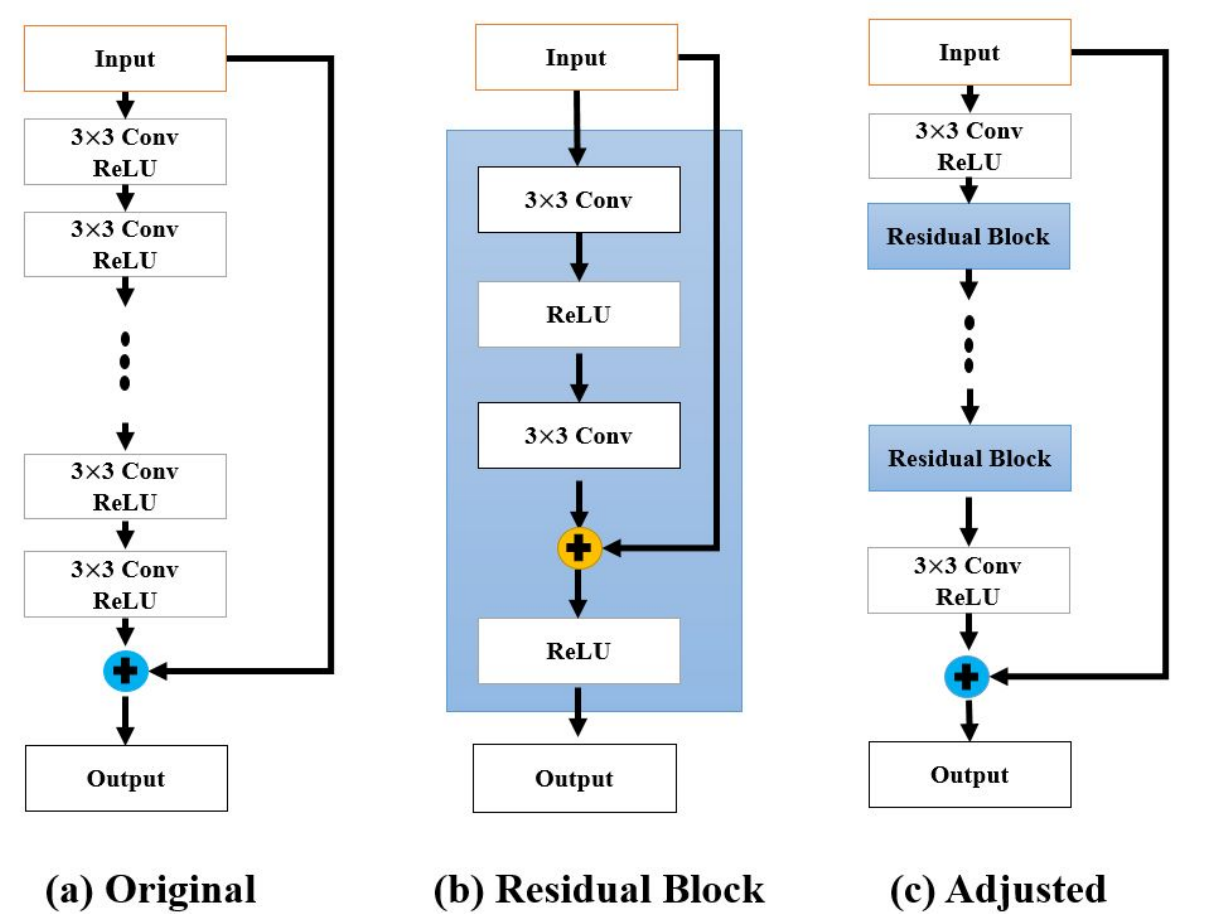}}
\caption{ The reconstruction module of DCCNN \cite{Schlemper2017ADC} and MoDL \cite{aggarwal2018modl}. A plain convolutional neural network with residual learning is adopted originally drawn in (a). In our re-implementation, we use residual blocks (b) instead of plain CNN drawn in (c) to avoid the gradient disappearance.}
\label{fig3}
\end{figure}

\begin{figure}[ht]
\centerline{\includegraphics[width=7.5cm]{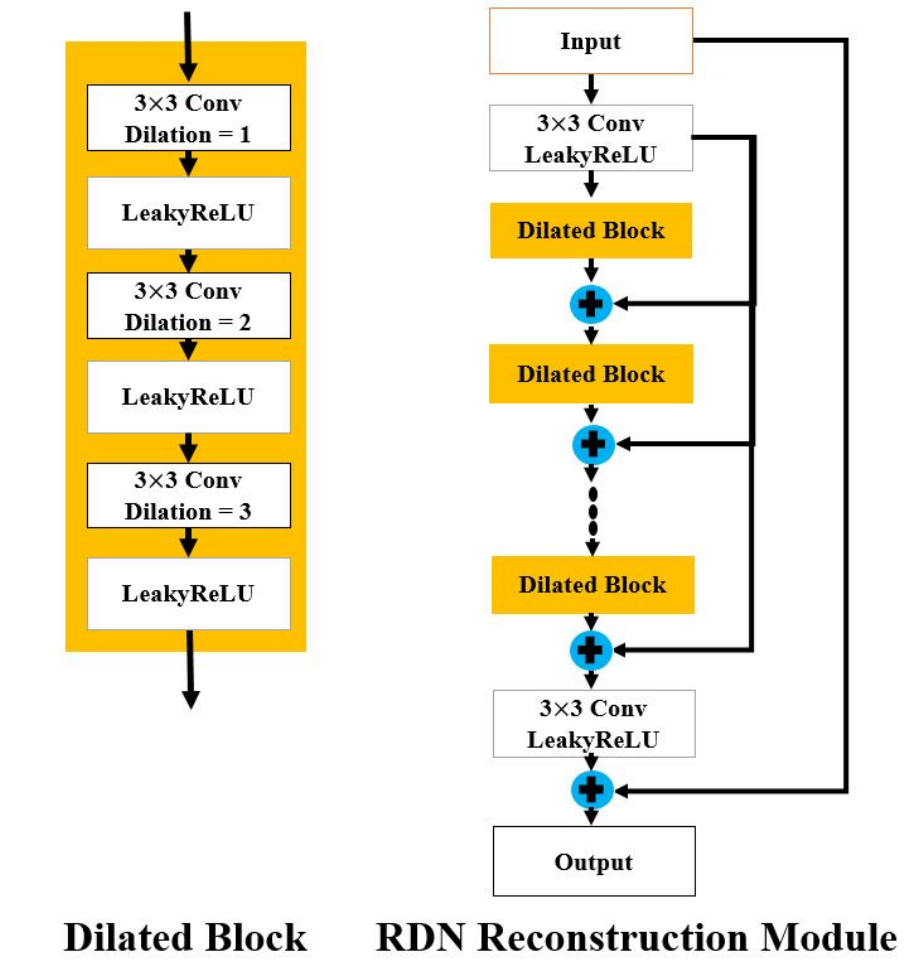}}
\caption{ The reconstruction module of RDN \cite{Sun2018CompressedSM}. The dilated convolution is used to expand perception field and improve the performance. All the yellow dilated blocks share the same weights in reconstruction module by recursive learning \cite{Kim2015DeeplyRecursiveCN} to reduce parameters. }
\label{fig4}
\end{figure}

Under this uniform framework, some previous works \cite{Huang2018MRIRV, Sun2018CompressedSM, Zeng2019AVD} discussed how the design of reconstruction modules can improve the quality of MR image reconstruction. As mentioned above, their networks were all built by handcraft. DCCNN\cite{Schlemper2017ADC} and MoDL \cite{aggarwal2018modl} all used a plain CNN as the reconstruction module with residual learning shown in Fig.\ref{fig3}. Particularly in MoDL\cite{aggarwal2018modl}, all the reconstruction modules shared the same weights to reduce the number of parameters. Sun et al.\cite{Sun2018CompressedSM} designed a recursive dilated network named RDN shown in Fig.\ref{fig4}.   

\begin{figure}[!h]
\centerline{\includegraphics[width=5cm]{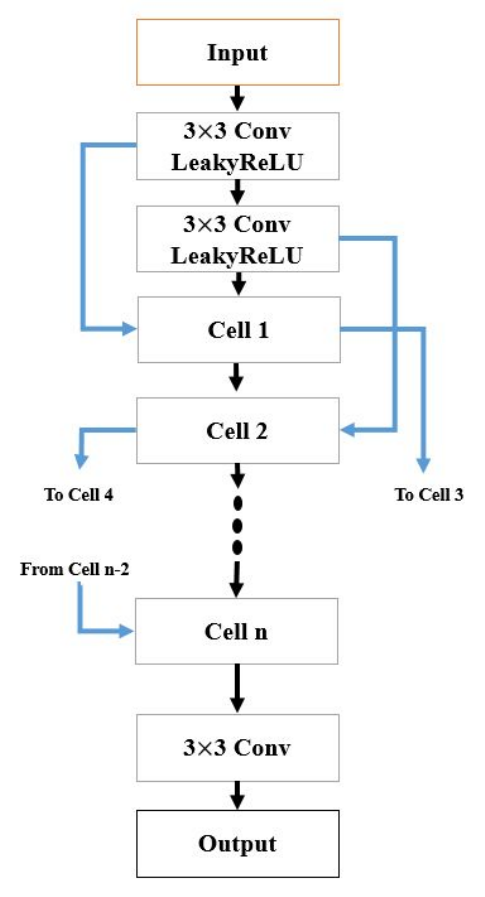}}
\caption{ The reconstruction module used in our network. The first and the last common $3 \times 3$ convolutional layers are used to refine the channels of the input and the output data. The searched cells are stacked to form the whole architecture.}
\label{fig5}
\end{figure}

In this work, we searched for the internal structure of the \emph{cells}, which are stacked to form the reconstruction module drawn in Fig.\ref{fig5} via the NAS algorithm automatically.
The concept of cell is also adopted in previous NAS works \cite{liu2018darts, liu2019auto, pham2018efficient} and will be introduced in the following section. The first and the last common $3 \times 3$ convolutional layers are used to refine the channels of the input and the output data similar to previous works \cite{Schlemper2017ADC, Sun2018CompressedSM}. 

\subsection{Differentiable Search Strategy}

\subsubsection{Definition of the Cell}
In this manuscript, a cell maps the output tensors of previous two cells to construct its output, i.e. we can have:

\begin{equation}
C_l = Cell \left( C_{l-1}, C_{l-2}, \alpha \right),
\end{equation}
where $C_l$ is the output of cell $l$ , $\alpha$ is a parameter representing the relaxation of discrete inner cell architectures by the following formulations.

The inner architecture of $C_l$ can be defined as a directed acyclic computation graph formed by sequential internal nodes $[P_l^1, P_l^2, \cdots, P_l^n]$ shown in Fig.\ref{fig5}, and we can have:

\begin{equation}
C_l = Concat( P_l^1, P_l^2, \cdots, P_l^n ).
\end{equation}
Define the connections between two nodes as a selection from candidate layer operations set $\mathcal{O}$, the input of $P_l^i$ as a selection from tensors set $\mathcal{I}_l^i$. $\mathcal{O}$ contains different CNN layer types, such as $3 \times 3$ convolutional layer. The latter node can take in the input of this cell and all previous nodes' outputs, so we have: $\mathcal{I}_l^1 = [C_{l-1}, C_{l-2}]$, $\mathcal{I}_l^2 = [C_{l-1}, C_{l-2}, P_l^1]$, $\mathcal{I}_l^3 = [C_{l-1}, C_{l-2}, P_l^1, P_l^2]$, $\cdots$.

\begin{figure*}
\centerline{\includegraphics[width=14cm]{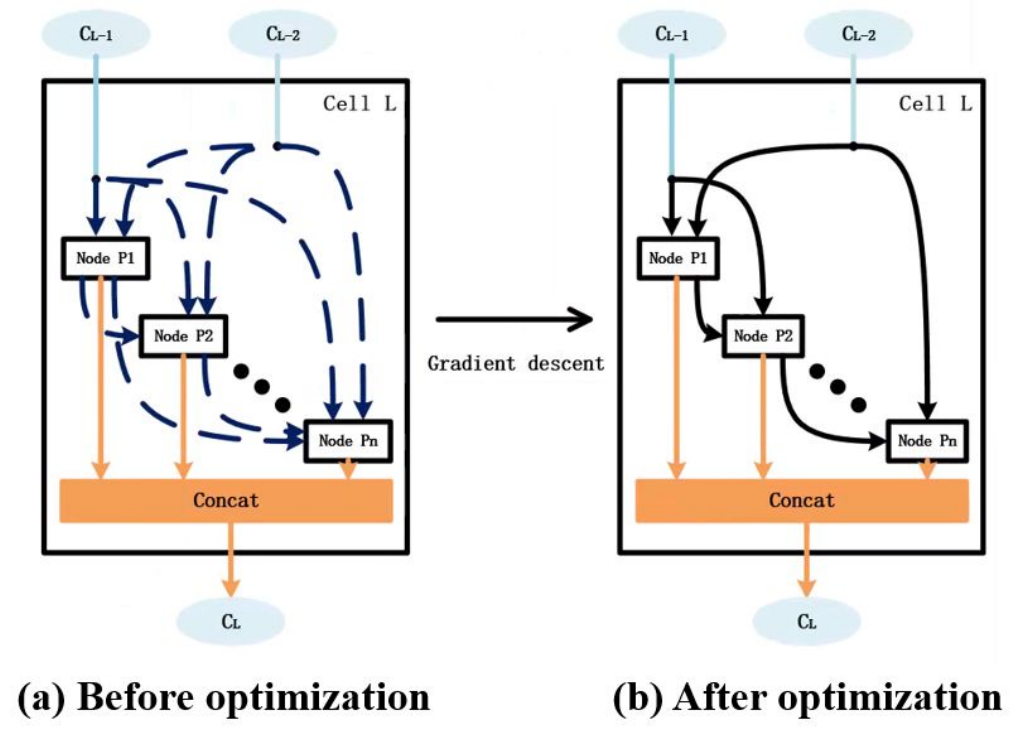}}
\caption{ The definition of the cell and parameter $\alpha$. A cell takes in the output tensors of the previous two cells to construct its own output. Internal nodes with connections formed the cell are defined as a directed acyclic computation graph. Before searching for the final architecture, these nodes are densely connected by all possible layer types shown in (a). After optimization, the best two connections with its most suitable layer type are preserved according to $\alpha$ and form the final architecture in (b).}
\label{fig6}
\end{figure*}

Before searching internal cell architecture, these nodes are densely connected by all possible layer types in $\mathcal{O}$ shown in Fig.\ref{fig6} (a). The parameter $\alpha$ is integrated into each cell node by the following two steps.

First, the output of $P_l^i$ is defined by tensors in $\mathcal{I}_l^i$, that is:
\begin{equation}
P_l^i = \sum_{ I_l^j \in \mathcal{I}_l^i } O_{ j \rightarrow i} (I_l^j),
\end{equation}

Second, the parameter $\alpha$ is added as the probability associated with each operation $O^k \in \mathcal{O}$ by:
\begin{equation}
O_{ j \rightarrow i} (I_l^j) = \sum_{ O^k \in \mathcal{O} } \alpha_{ j \rightarrow i}^k O^k (I_l^j),
\end{equation}
where $\alpha$ is limited by:

\begin{equation}
\sum_{ k=1}^{|\mathcal{O}|} \alpha_{ j \rightarrow i}^k = 1,    \forall i > j,
\end{equation}
\begin{equation}
\alpha_{ j \rightarrow i}^k \ge 0, \forall i > j, O^k \in \mathcal{O}.
\end{equation}

With the introduction of $\alpha$, the cell architecture search problem can be successfully integrated into a differentiable computation graph. After optimizing $\alpha$ via gradient descent, the layer operations with top-2 $\alpha$-value are preserved to form the final structure shown in Fig.\ref{fig6} (b).

To clarify the final searched cell structure,  we can define a 4-tuple $ P_l^i = [ I_{l}^{i_1}, I_{l}^{i_2}, O_{l}^{i_1}, O_{l}^{i_2} ]$, where $I_{l}^{i_1}, I_{l}^{i_2} \in \mathcal{I}_{l}^{i}$ are selections of the input tensors and $O_{l}^{i_1}, O_{l}^{i_2} \in \mathcal{O}$ are selections of candidate layer operations based on the optimized $\alpha$ value. Thus we have:

\begin{equation}
P_l^i = O_l^{i_1}(I_l^{i_1})+ O_l^{i_2}(I_l^{i_2}).
\end{equation}
Finally, the structure of the cell is defined and can then be used as a common CNN module. 

\subsubsection{Design of the Operation Search Space}

\begin{figure}[!h]
\centerline{\includegraphics[width=9cm]{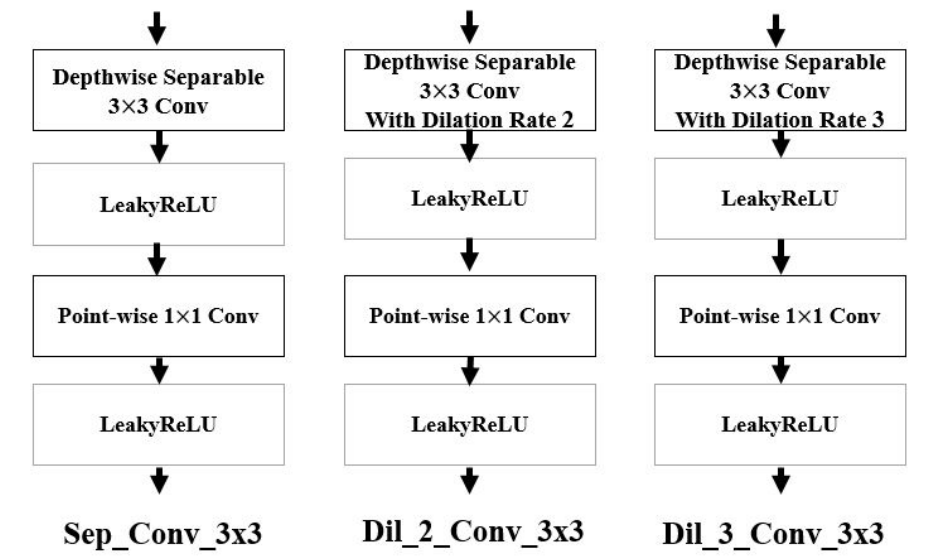}}
\caption{ The candidate layer operations set $\mathcal{O}$. }
\label{fig7}
\end{figure}
 
 The candidate layer operations set $\mathcal{O}$ was defined by us as follows shown in Fig.\ref{fig7}:
\begin{itemize}
    \item Sep\_Conv\_3x3 : The $3 \times 3$ separable convolutional layer is formed by cascading a $3 \times 3$ depthwise separable convolutional layer and a $1 \times 1$ pointwise layer.
    \item Dil\_2\_Conv\_3x3 : The $3 \times 3$ separable convolutional layer with a dilation rate of 2 in the depthwise separable layer.
    \item Dil\_3\_Conv\_3x3 : The $3 \times 3$ separable convolutional layer with a dilation rate of 3 in the depthwise separable layer.
    \item Skip connect
    \item None connect
\end{itemize}

We chose these operations based on the following observation and summary of previous works: 
\begin{itemize}
    \item The separable convolution was proposed by \cite{chollet2017xception} and  widely used in Mobilenet\cite{howard2017mobilenets}, Shufflenet\cite{zhang2018shufflenet} and other efficient networks. Comparing with common convolutional layer, separable convolution can use fewer calculations and parameters.
    \item The dilated convolution \cite{Yu2015MultiScaleCA} is widely adopted in low-level image analysis tasks, because it can expand perception field without adding parameters. RDN \cite{Sun2018CompressedSM} shows that dilated layers with various dilation rates CAN benefit the performance of MR image reconstruction.
\end{itemize}

\subsubsection{Optimization Strategy}

We followed the optimization strategy in \cite{liu2018darts} to search the inner structure of cells. The training data was divided into two disjoint sets train\_$\omega$ and train\_$\alpha$ according to the first-order approximation. The disjoint set partition also prevents the architecture from over-fitting the whole training data. 

The optimization alternates between:

1. Update network weights $\omega$ by $\nabla_\omega \mathcal{L}_{train\_\omega} (\omega,\alpha) $,

2. Update architecture $\alpha$ by $\nabla_\alpha \mathcal{L}_{train\_\alpha} (\omega,\alpha) $.

The optimization object function $\mathcal{L} (\omega,\alpha) $ is defined by integrating $\alpha$ to Eq.\ref{equ:5} as:

\begin{equation}
\mathop{{\min}}\limits_{{\omega}} \mathcal{L} (\mathcal{D}(x, \omega, \alpha)-y).
\end{equation}

According to \cite{Zhao2017Loss}, the L1 loss is beneficial to train machine learning models on computer vision tasks, even when the evaluation is performed under L2 norm related metrics, e.g. PSNR. Inspired by this, we defined the loss function $\mathcal{L}$ as L1 loss between paired zero-filled MR image $x$ and fully-sampled MR image $y$.  The searching process needs to be stopped when the cell structure starts to keep stable according to early stopping strategy, which is commonly used in NAS works \cite{fiszelew2007finding, baker2017accelerating}. 

After the structure of the whole network is determined, the final network needs to be re-trained on the whole training set to maximize its final reconstruction performance.

%\subsubsection{Complexity analysis}

%To some degree, we sampled a smaller but most important architecture from a densely connected large network according to optimized $\alpha$. Given a pre-searched architecture with N cells and M internal nodes for each and operation search space contains J types, we can use the number of total operations in this computation graph to approximate computation complexity. The complexity reduced can be approximated as:

%\begin{equation}
%N \times \frac{(M+3)M}{2} \times  J \rightarrow N \times 2M \times 1.
%\end{equation}

\section{Experimental Analysis}

We compared our framework with the following methods: the conventional Total Variation (TV) \cite{Rudin1992NonlinearTV} minimization based iterative
 algorithm, U-net baseline model used in \cite{fastMRI} as a  typically data-driven method, DCCNN \cite{Schlemper2017ADC}, MoDL \cite{aggarwal2018modl}, and RDN \cite{Sun2018CompressedSM} which can be regarded as a representative improved approach following \cite{Schlemper2017ADC, aggarwal2018modl}.

\subsection{Dataset and Pre-processing}

We conducted all the experiments on the fastMRI dataset \cite{fastMRI} including knee and brain subsets obtained on 3 and 1.5 Tesla magnets. Due to limited computation resources we have, we randomly selected 80 scans with 2829 slices as the training set and 40 scans with 1457 slices as the testing set from the knee subset, and this mini-fastMRI knee dataset contains more slices than experiments in DCCNN\cite{Schlemper2017ADC} and RDN\cite{Sun2018CompressedSM}. To evaluate the generalizability of the searched architecture for different organ reconstruction tasks, extensive experiments were carried out on a mini-fastMRI brain dataset including 100 scans with 1600 slices as the training set and 50 scans with 800 slices as the testing set from the T2 weighted brain subset.

The mini-fastMRI knee dataset provides raw complex-valued fully-sampled single-coil k-space data with different sizes. The following steps were performed to make paired $320 \times 320$ zero-filled and fully-sampled reconstructions: First, 2D-IFFT was applied on original k-space data to get MR images, which were then cropped centrally to generate $320 \times 320$ cropped complex images. After that, 2D-FFT was performed on each and obtain the corresponding $320 \times 320$ fully-sampled k-space data. The mini-fastMRI brain dataset, however, does not offer the complex-valued single-coil k-space data but has real-valued brain MR images with the same size of $320 \times 320$. We treated the real-valued brain MR images as complex-valued with zero phases following \cite{Sun2018CompressedSM, Zeng2019AVD} so that 2D-FFT could be applied to get simulate complex-valued k-space data. 

To simulate the accelerated signals acquisition process, we used 4-fold and 8-fold Cartesian sub-sampling masks following the settings in \cite{fastMRI}: For 4-fold sub-sampling, the central $8\%$ k-space lines are full-sampled with the outer k-space under-sampled uniformly. For 8-fold sub-sampling, the central $4\%$ k-space lines are full-sampled with the outer k-space under-sampled uniformly. 

All the MR images were normalized by magnitude to [-6,6] with their phases unchanged. After normalization, the complex-valued images were separated into the real and the imaginary image and concatenated as the 2-channel input of the reconstruction network.

\subsection{Implementation Details}

\subsubsection{Baseline configurations}

The BART toolkit \cite{BART} was used to perform the TV minimization iterative reconstruction algorithm. We set the total variation regularization weight as 0.01 and implemented 200 iterations on each slice.

All the deep CNNs in this manuscript were implemented with Pytorch \cite{paszke2019pytorch}. The input and output of all the CNNs are 2 channels and all the other layers have 32-channel output except for special reminders (e.g. U-net). No normalization operation, e.g. batch-normalization \cite{Ioffe2015BatchNA},  was used in these networks following DCCNN \cite{Schlemper2017ADC}.

\begin{figure}[!h]
\centerline{\includegraphics[width=\columnwidth]{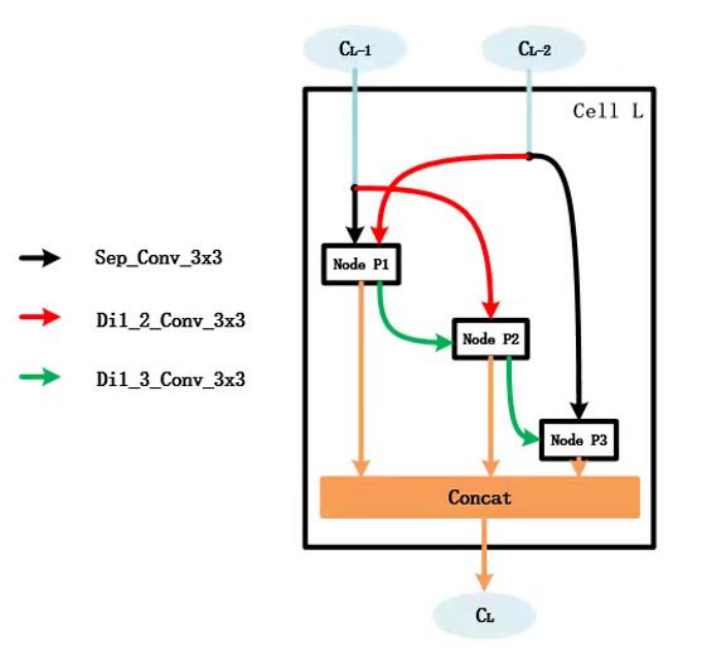}}
\caption{ The final searched cell structure with 3 internal nodes, named NasN\_Knee. Two input connections with their independent layer type are preserved to form the whole architecture.}
\label{fig8}
\end{figure}

We searched for our reconstruction module on the knee dataset with 3 cascading cells and each cell includes 3 internal computation nodes, named NasN\_Knee. The final searched cell structure is drawn in Fig.\ref{fig8}. The outputs of $C_{l-1}$ and $C_{l-2}$ are processed by $1 \times 1$ convolutional layers in $C_{l}$ to reduce channel number from 96 to 32.

To make different networks comparable, we made different networks have relatively similar learn-able parameters and floating-point operations (FLOPs).  As a result, the reconstruction module of DCCNN\cite{Schlemper2017ADC}  contains 3 residual blocks, and the reconstruction module of RDN \cite{Sun2018CompressedSM} contains 3 recursive dilated blocks. Because we are dealing with single-coil reconstruction, we re-trained the DCCNN with sharing-weight reconstruction modules to re-implement MoDL \cite{aggarwal2018modl}. For U-net, the input data are down-sampled 4 times with channels doubled starting from 32 channels. 

According to DCCNN \cite{Schlemper2017ADC} and RDN \cite{Sun2018CompressedSM}, more blocks in their reconstruction modules will produce better results. Thus, we also added more blocks in these networks to evaluate whether they can use more computation resources to achieve similar performance with our searched network.

\subsubsection{Training Strategy}

All deep networks were trained on one TITAN X Pascal GPU with 12GB memory using Adam optimizer \cite{kingma2014adam} for parameter learning with L1 loss only and a batchsize of 2. The initial learning rate was set to be 0.001 for the first 40 epochs and 0.0001 for the later 40 epochs. During the training process, the 4-fold and 8-fold Cartesian sub-sampling mask was generated randomly for every slice with equal possibility. This can also be viewed as a data augmentation to avoid over-fitting \cite{Caruana2000OverfittingIN}.

\subsubsection{Evaluation Strategy}

For all deep methods, the FLOPs per inference were calculated by counting multiplication and add operations in all convolutional layers inside the reconstruction modules after feeding a $320 \times 320$ 2-channel MR image. The FLOPs is a wildly used metric to evaluate the computation resource cost in previous NAS works \cite{liu2018darts, liu2019auto}. And the numbers of learn-able parameters are also provided, because previous works \cite{aggarwal2018modl, Sun2018CompressedSM} claimed they used fewer parameters to achieve better performance.

We evaluated the MSE, Normalized MSE (NMSE), PSNR and SSIM \cite{Zhou2004Image} between reconstructed results and target fully-sampling MR images on magnitude with the similar setting in baseline works \cite{aggarwal2018modl,Schlemper2017ADC, Sun2018CompressedSM}. In evaluation, half of the total cases were sub-sampled with 4-fold Cartesian sub-sampling masks and the other with 8-fold ones. All quantitative evaluation results were calculated on images reconstructed from the same corresponding sub-sampling patterns.

\subsection{Knee MR Reconstruction Comparison}

\begin{table*}[!t]
\caption{\label{tab1}Quantitative knee MR reconstruction results of different methods. The results are calculated on the testing set including 40 scans. ($AVG \pm STD$)}
\centering
\begin{tabular}{p{4cm}|p{2cm}|p{2cm}|p{2cm}|p{2cm}|p{1cm}|p{1cm}}
\hline
Model & MSE ($\times$1e-10)       & NMSE($\times$1e-2)     & PSNR(dB)    & SSIM    & FLOPs & Param. \\
\hline
TV\cite{Rudin1992NonlinearTV}     & $2.351 \pm 5.355$           & $7.104 \pm 7.633$                 & $28.12 \pm 6.246$                    & $0.5192 \pm 0.2509$               &  N/A                   &  N/A                    \\
U-net\cite{fastMRI}     & $1.764 \pm 3.742$           & $5.654 \pm 6.321$                 & $29.15 \pm 5.904$                    & $0.6028 \pm 0.2508$               & 12.17G                   & 3349K                             \\
DCCNN\cite{Schlemper2017ADC} with 3 residual blocks                                                                 & $1.718 \pm 4.021$               & $5.521 \pm 6.484$             & $29.37 \pm 6.538$                    & $0.6118 \pm 0.2648$                   & 17.41G                   & 170.0K
                           \\
DCCNN\cite{Schlemper2017ADC} with 11 residual blocks                                                                & $1.635 \pm 3.651 $                & $5.407 \pm 6.456$                  & $29.49 \pm 6.544$                    & $0.6147 \pm 0.2658$                  & 62.87G                   & 613.9K                            \\
MoDL\cite{aggarwal2018modl} with 3 residual blocks                                                                 & $2.025 \pm 5.167$               & $5.886 \pm 6.358$             & $28.95 \pm 6.226$                    & $0.6075 \pm 0.2596$                   & 17.41G                   & 56.7K
                           \\
MoDL\cite{aggarwal2018modl} with 11 residual blocks                                                                & $1.938 \pm 4.785 $                & $5.788 \pm 6.356$                  & $29.04 \pm 6.235$                    & $0.6089 \pm 0.2606$                  & 62.87G                   & 204.3K                            \\
RDN\cite{Sun2018CompressedSM} with 3 recursive dilated blocks                                                     & $1.623 \pm 3.587 $                & $5.398 \pm 6.444$                  & $29.49 \pm 6.501$                    & $0.6147 \pm 0.2656$                   & 26.03G     & 86.79K
                            \\
RDN\cite{Sun2018CompressedSM} with 8 recursive dilated blocks                                                   & $1.531 \pm 3.204 $                & $5.290 \pm 6.468$                  & $29.62 \pm 6.534$                    & $0.6170 \pm 0.2665$                   & 68.79G                   & 86.79K                            \\
                            
NasN\_Knee                                                                  & $1.432 \pm 2.919$       & $5.112 \pm 6.408$        & $29.83 \pm 6.692$            & $ 0.6204 \pm 0.2676$                   & 15.07G
       & 142.3K \\
\hline
\end{tabular}
\end{table*}

The quantitative evaluation results of knee MR reconstruction are shown in Tab.\ref{tab1}, demonstrating that searched NasN\_Knee network outperforms previous state-of-the-art frameworks. Among all the deep models, U-net is more different because its feature maps are reduced in size after down-sampling with channels doubled. So U-net has much more learn-able parameters than others but with fewer operations. U-net has $40\times$ more parameters than our proposed network while fails to provide better results. DCCNN, MoDL, RDN, and our NasN\_Knee all adopt k-space fusion strategy. Comparing to DCCNN, MoDL uses sharing-weight reconstruction modules to reduce parameters and leads to worse reconstruction results. Because dilated blocks in RDN share the same weights, so its learn-able parameters do not increase with more blocks. Although RDN and MoDL contain fewer learn-able parameters, ``there is no free lunch", the FLOPs do not decrease by recursive learning, i.e. the inference speed is still limited. Our searched NasN\_Knee architecture uses $4\times$ fewer FLOPs to reconstruct better results than RDN with 8 recursive dilated blocks.

\begin{figure*}
\centerline{\includegraphics[width=18cm]{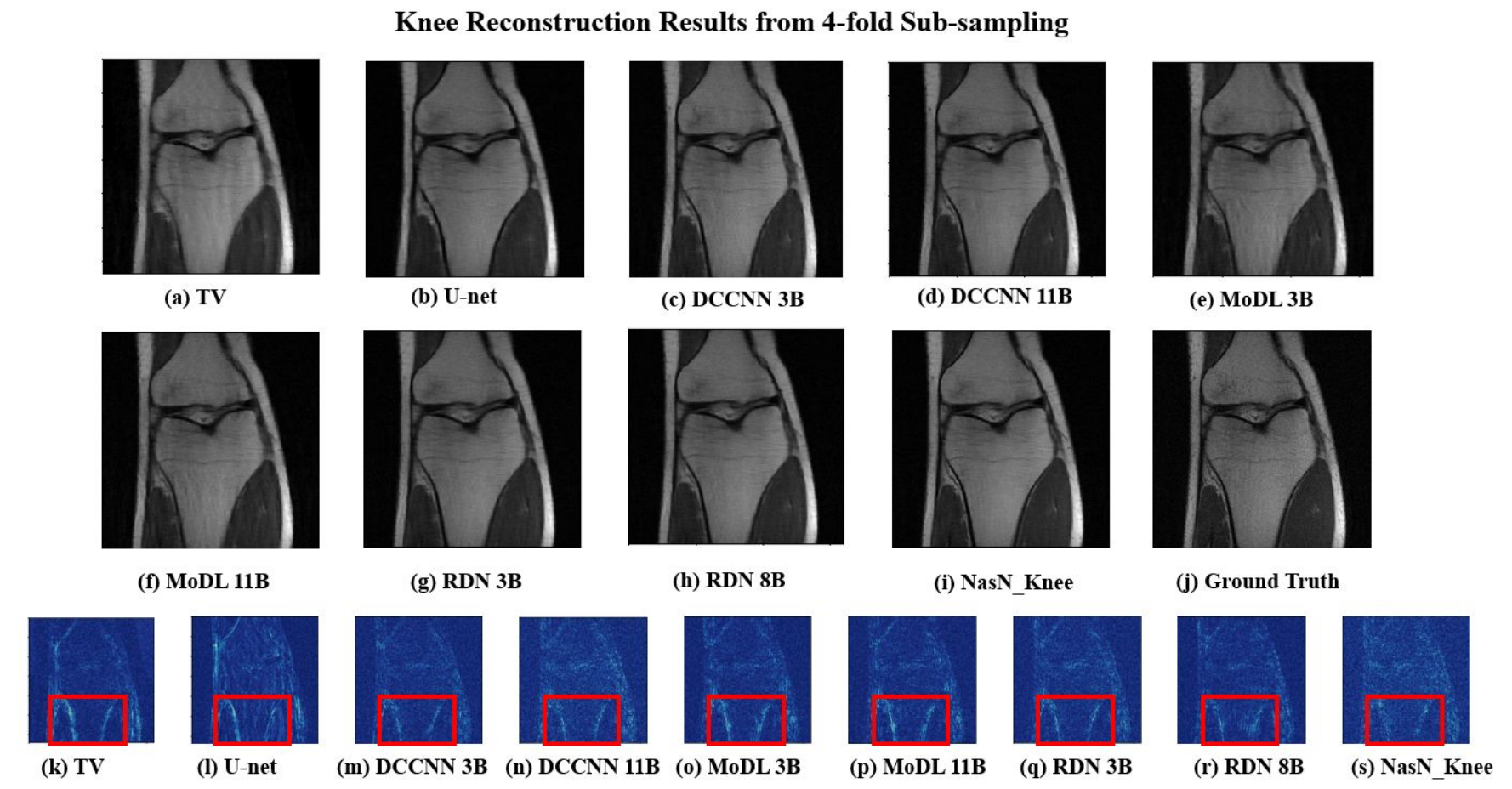}}
\caption{The qualitative knee MR reconstruction results of different methods with 4-fold sub-sampling. Here the red boxes address where our reconstruction results have less noise.}
\label{fig9}
\end{figure*}

\begin{figure*}
\centerline{\includegraphics[width=18cm]{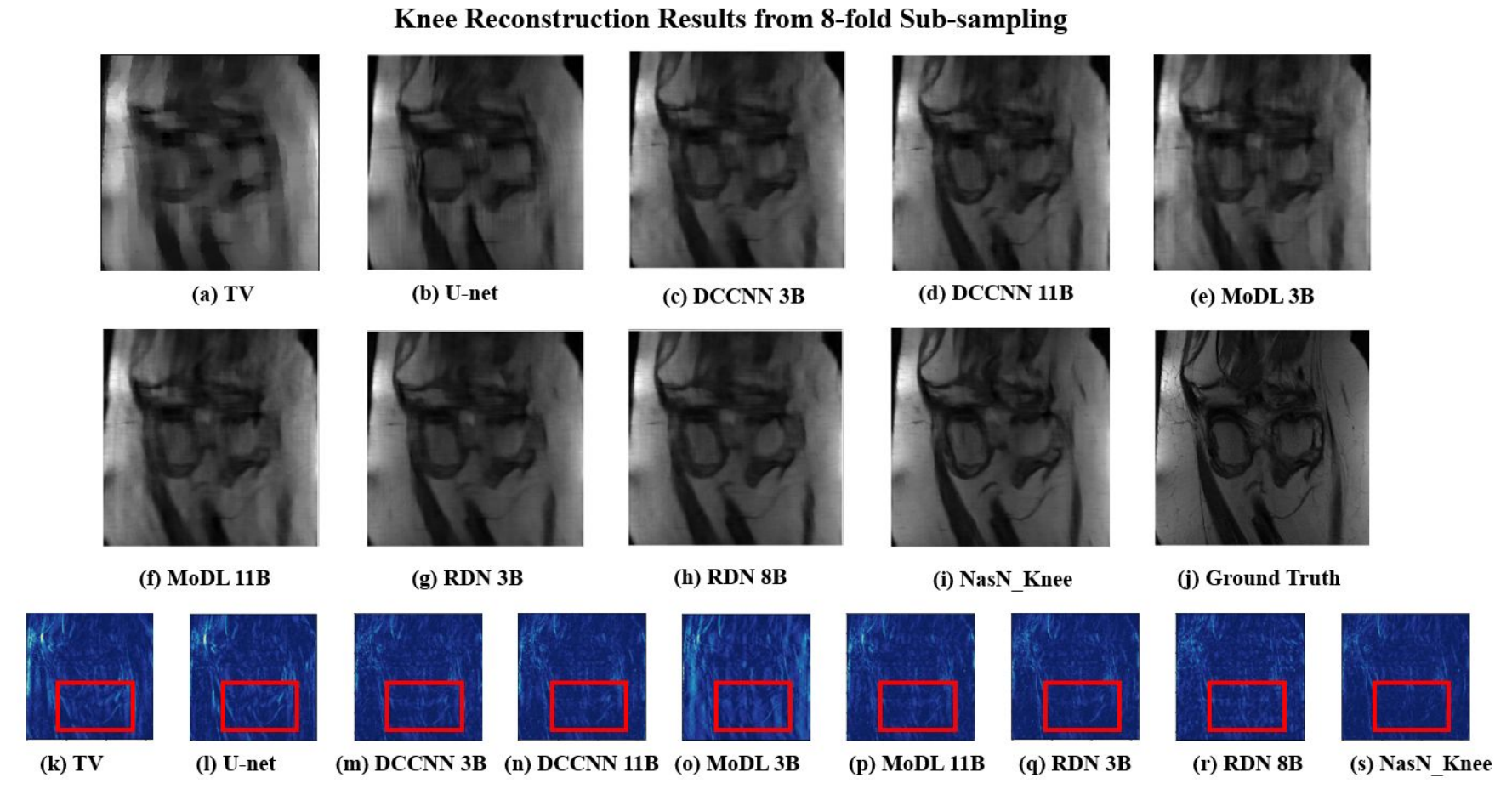}}
\caption{The qualitative knee MR reconstruction results of different methods with 8-fold sub-sampling. Here the red boxes address where our reconstruction results have less noise.}
\label{fig10}
\end{figure*}

The qualitative knee MR reconstruction results are shown in Fig.\ref{fig9} and Fig.\ref{fig10}. Two slices reconstructed from different sub-sampling ratios are presented, where we can find NasN\_Knee can reduce aliasing artifacts more effectively compared with other methods. When the sub-sampling ratio gets bigger, our model performs much better than other methods with less noise and more accurate structural details.

\subsection{Analysis of Hyper-Parameters}
Although the architecture of cells is searched automatically, there still exist some hyper-parameters set manually in the whole framework. In this part, we conduct experiments to explore how these hyper-parameters may affect the reconstruction performance and the searched architecture.

\subsubsection{Number of Cascading Modules and Cells}
Given the searched cell structure (shown in Fig.\ref{fig8}), we can cascade more modules (shown in Fig.\ref{fig2}) or more cells in each module (shown in Fig.\ref{fig5}) to build the whole reconstruction framework. 

\begin{figure}[!h]
\centerline{\includegraphics[width=\columnwidth]{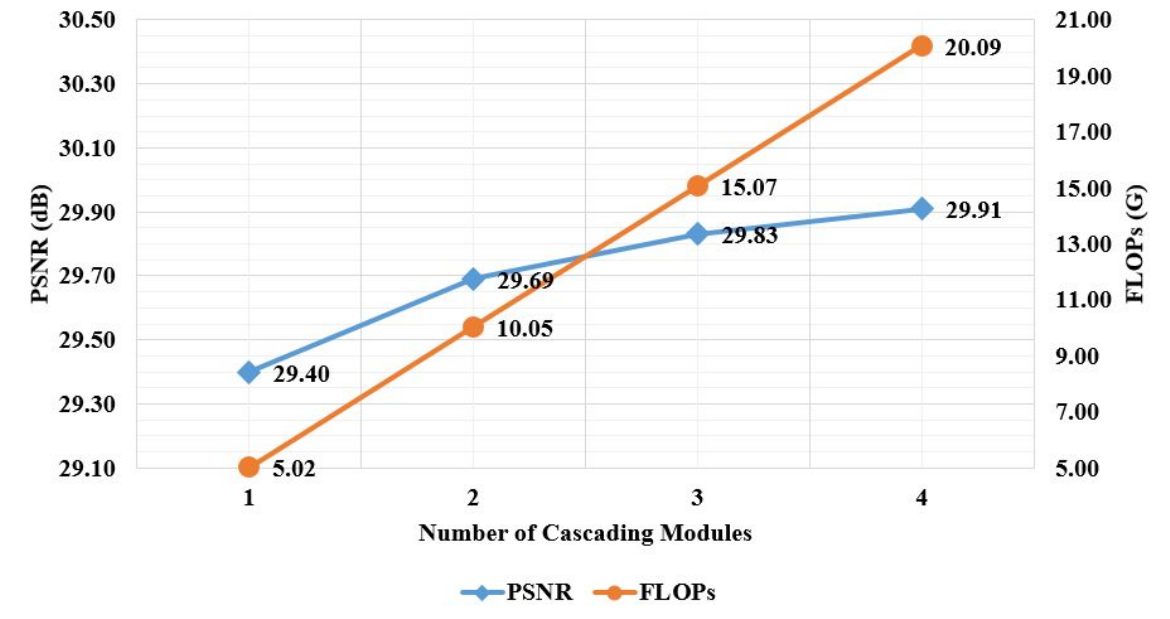}}
\caption{ The reconstruction performance and FLOPs as functions of cascading modules.}
\label{fig11}
\end{figure}

\begin{figure}[!h]
\centerline{\includegraphics[width=\columnwidth]{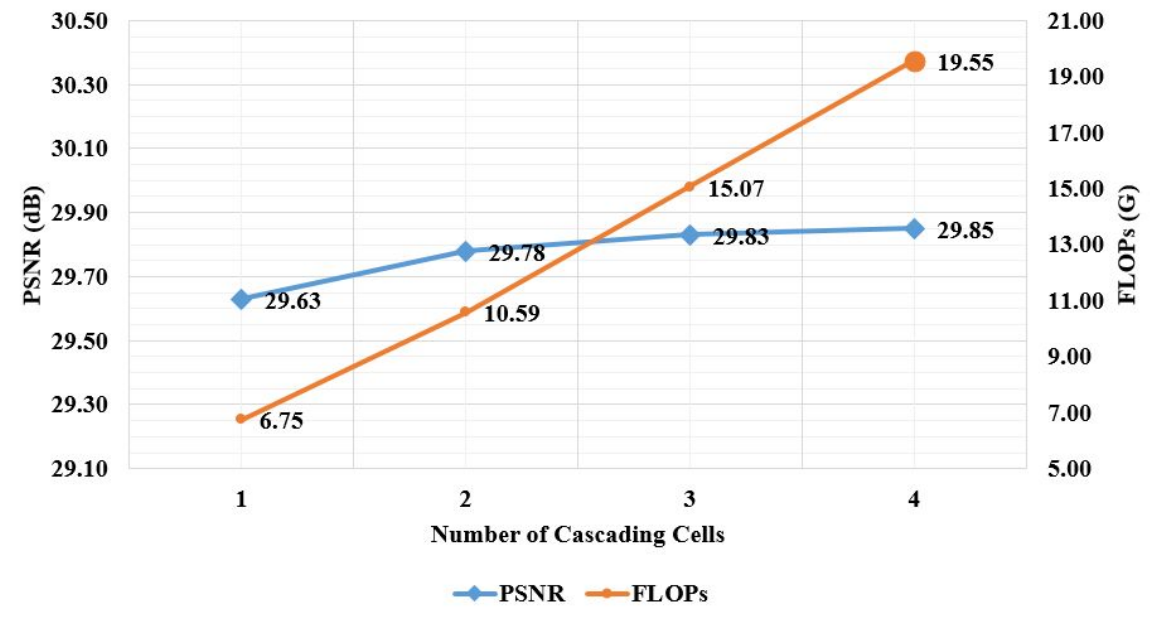}}
\caption{ The reconstruction performance and FLOPs as functions of cascading cells in each module.}
\label{fig12}
\end{figure}

The reconstruction performance and FLOPs as functions of cascading modules are shown in Fig.\ref{fig11}. While the reconstruction performance and FLOPs as functions of cascading cells in each module are shown in Fig.\ref{fig12}. 

 We can find that more cascades lead to better reconstruction results but with heavier computation load. Comparing to the increase of FLOPs, the improve of performance may be not efficient enough.

\subsubsection{Number of Internal Nodes in Each Cell}
\begin{figure*}
\centerline{\includegraphics[width=14cm]{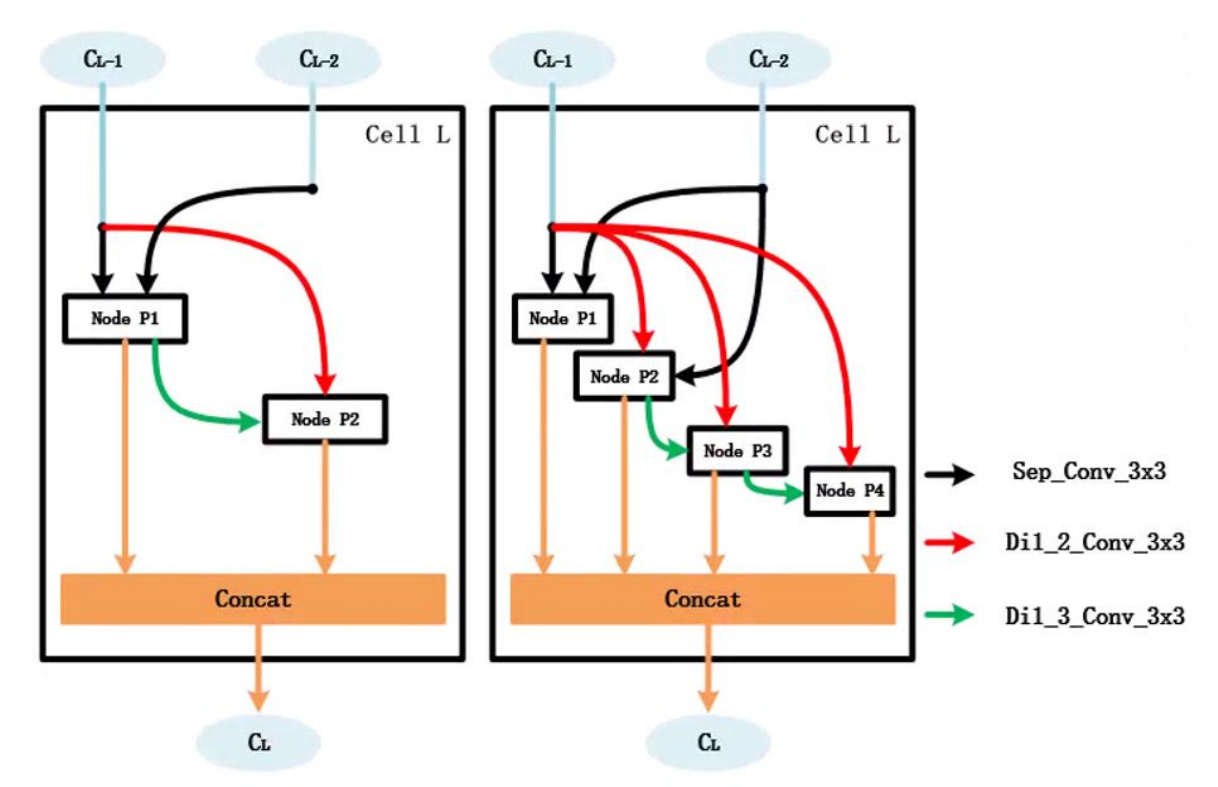}}
\caption{ The searched structures of the cell with 2 and 4 nodes inside. Convolutional layers with larger perception field are always placed in the bottom and Feature maps from different levels are fused properly. By observing the selection of connections and layer types, some insights of network design can be concluded.}
\label{fig13}
\end{figure*}

To explore the impact of internal nodes in each cell, we searched for different architectures where each cell contains a different number of internal nodes on the knee MR dataset. The searched results with 2 and 4 internal nodes are drawn in Fig.\ref{fig13}. The reconstruction performance of these architectures on the knee dataset is listed in Tab.\ref{tab2}. 

\begin{table*}[!t]
\caption{\label{tab2}Quantitative evaluation results of searched architectures with different internal nodes. The results are calculated on the testing set including 40 scans. ($AVG \pm STD$)}
\centering
\begin{tabular}{p{2.5cm}|p{2cm}|p{2cm}|p{2cm}|p{2cm}|p{1cm}|p{1cm}}
\hline
Number of Nodes & MSE ($\times$1e-10)       & NMSE($\times$1e-2)    & PSNR(dB)    & SSIM    & FLOPs  & Param. \\
\hline
2      &  $1.462 \pm 3.046$               & $5.148 \pm 6.407$                  &  $29.78 \pm 6.696$                   & $0.6201 \pm 0.2684$               & 11.37G                   & 107.7K                             \\
3      & $1.432 \pm 2.919$       & $5.112 \pm 6.408$        & $29.83 \pm 6.692$            & $ 0.6204 \pm 0.2676$                   & 15.07G
       & 142.3K \\
4                                                                & $1.405 \pm 2.781$               & $5.081 \pm 6.420$                  & $29.87 \pm 6.719$ 
&  $0.6218 \pm 0.2687$                   & 18.76G                   & 176.8K                            \\
\hline
\end{tabular}
\end{table*}

The results show that the increasing internal nodes bring mild but consistent benefit for the final performance. With 2 internal nodes and 6 times fewer FLOPs, the searched architecture still outperforms RDN\cite{Sun2018CompressedSM}. And the found structures shown in Fig.\ref{fig8} and Fig.\ref{fig10} demonstrate that the algorithm does learn some principles to form the reconstruction architecture automatically. These principles offer insights for other researchers to design their network for other MR medical image tasks, e.g. Semantic Segmentation \cite{danelakis2018survey} or Super-Resolution \cite{zhao2020super}:

\begin{itemize}
    \item Deeper feature maps require larger perception fields. In our searched results, layers with dilation rate 3 are always preferred in the bottom of the cell behind layers with dilation rate 2.
    \item It is important to fuse feature maps from different levels properly. In our searched cell structure, feature maps from $C_{l-2}$ are used to refine different nodes.
\end{itemize}

\subsubsection{The Operation Search Space}
To appreciate the value of the operation search space, experiments were performed to explore the relationship between operation search space and network performance.

Note original operation search space by A, we define a new space B as follows:
\begin{itemize}
    \item Sep\_Conv\_3x3
    \item Dil\_2\_Conv\_3x3
    \item Conv\_9x1\_1x9 : two cascading depthwise separable convolutional layers with the kernel size of $9 \times 1$ and $1 \times 9$.
    \item Skip Connect
    \item None Connect
\end{itemize}

\begin{figure}[!h]
\centerline{\includegraphics[width=3cm]{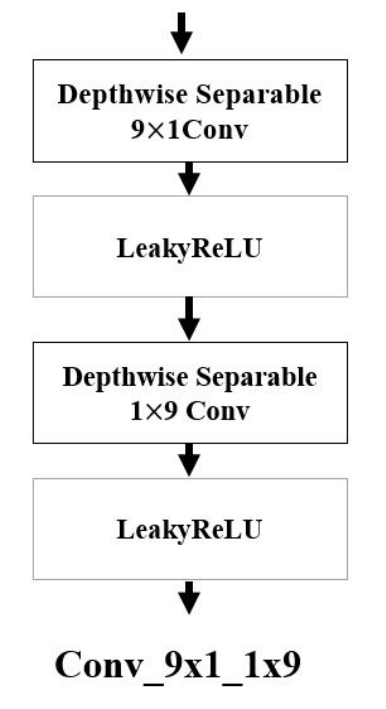}}
\caption{ The structure of Conv\_9x1\_1x9. Two cascading convolutional layers with $9 \times 1$ and $1 \times 9$ kernel size are adopted to achieve a $9 \times 9$ perception field.}
\label{fig14}
\end{figure}

The design of Conv\_9x1\_1x9 shown in Fig.\ref{fig14} is inspired by \cite{peng2017large}, where large convolutional kernels lead to better performance in semantic segmentation. Two cascading $9 \times 1$ and $1 \times 9$ convolutional layers enable dense connections within a large $9 \times 9$ region when producing the feature map.

\begin{figure}[!h]
\centerline{\includegraphics[width=\columnwidth]{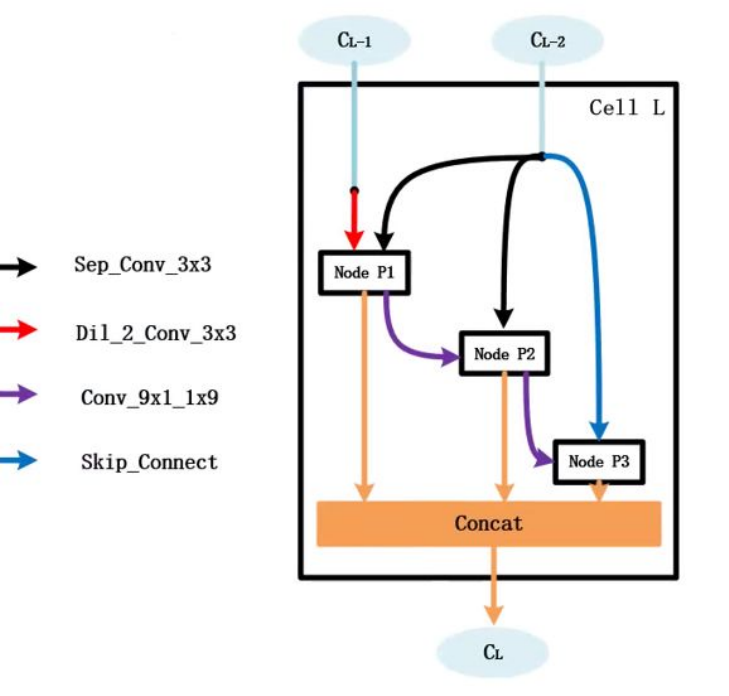}}
\caption{ The searched cell structure when the operation search space is redefined.}
\label{fig15}
\end{figure}

The search result is drawn in Fig.\ref{fig15} and the reconstruction results of these architectures are listed in Tab.\ref{tab3}. Results show that competitive reconstruction performance is achieved with fewer computation resources. With a different search space,  the principle that larger perception layers are placed in the bottom still holds. What's more, this experiment shows that the setting of operation search space allows us to explore more possibilities, if there are some novel and efficient layer types proposed in the future.

\begin{table*}[!t]
\caption{\label{tab3}Quantitative evaluation results of searched architectures with different operation search spaces. The results are calculated on the testing set including 40 cases. ($AVG \pm STD$)}
\centering
\begin{tabular}{p{3cm}|p{2cm}|p{2cm}|p{2cm}|p{2cm}|p{1cm}|p{1cm}}
\hline
Operation Search Space & MSE ($\times$1e-10)       & NMSE ($\times$1e-2)     & PSNR(dB)    & SSIM    & FLOPs & Param. \\
\hline
A      & $1.432 \pm 2.919$       & $5.112 \pm 6.408$        & $29.83 \pm 6.692$            & $ 0.6204 \pm 0.2676$                   & 15.07G
       & 142.3K \\
B      & $1.439 \pm 2.935$       & $5.122 \pm 6.401$        & $29.81 \pm 6.693$            & $ 0.6202 \pm 0.2678$                   & 12.44G
       & 117.2K \\
\hline
\end{tabular}
\end{table*}

\subsection{Generalizability of the Searched Architecture}

\begin{table*}[!t]
\caption{\label{tab4}Quantitative brain MR reconstruction results of different DL based methods. The results are calculated on the testing set including 50 scans. ($AVG \pm STD$)}
\centering
\begin{tabular}{p{3.5cm}|p{2cm}|p{2cm}|p{2cm}|p{2.5cm}|p{1cm}|p{1cm}}
\hline
Model & MSE ($\times$1e-10)       & NMSE($\times$1e-2)     & PSNR(dB)    & SSIM    & FLOPs & Param. \\
\hline
TV\cite{Rudin1992NonlinearTV}     & $16.81 \pm 16.39$           & $7.700 \pm 6.563$                 & $26.94 \pm 4.37$                    & $0.7594 \pm 0.1039$               & N/A                   & N/A                    \\
U-net\cite{fastMRI}     & $6.764 \pm 7.274$           & $3.055 \pm 2.697$                 & $31.00 \pm 4.764$                    & $0.8941 \pm 0.05672$               & 12.17G                   & 3349K                             \\
DCCNN\cite{Schlemper2017ADC} with 3 residual blocks                                                                 & $6.388 \pm 9.175$               & $2.894 \pm 3.833$             & $31.85 \pm 6.806$                    & $0.8933 \pm 0.07821$                   & 17.41G                   & 170.0K
                           \\
DCCNN\cite{Schlemper2017ADC} with 11 residual blocks                                                                & $5.609 \pm 7.532 $                & $2.534 \pm 3.051$                  & $32.25 \pm 6.332$                    & $0.8976 \pm 0.07016$                  & 62.87G                   & 613.9K                            \\
MoDL\cite{aggarwal2018modl} with 3 residual blocks                                                                 & $10.04 \pm 12.88$               & $4.561 \pm 5.326$             & $29.63 \pm 5.976$                    & $0.8558 \pm 0.09167$                   & 17.41G                   & 56.7K
                           \\
MoDL\cite{aggarwal2018modl} with 11 residual blocks                                                                & $17.98 \pm 16.98 $                & $8.208 \pm 6.489$                  & $26.60 \pm 4.107$                    & $0.7537 \pm 0.09623$                  & 62.87G                   & 204.3K                            \\
RDN\cite{Sun2018CompressedSM} with 3 recursive dilated blocks                                                     & $5.629 \pm 7.392 $                & $2.545 \pm 2.972$                  & $32.17 \pm 6.115$                    & $0.8957 \pm 0.06916$                   & 26.03G     & 86.79K
                            \\
RDN\cite{Sun2018CompressedSM} with 8 recursive dilated blocks                                                   & $5.782 \pm 7.440$                & $2.629 \pm 3.113$                  & $32.03 \pm 6.052$                    & $0.8944 \pm 0.06882$                   & 68.79G                   & 86.79K                            \\
                            
NasN\_Knee                                                                  & $4.596 \pm 6.236$       & $2.082 \pm 2.561$        & $33.12 \pm 6.348$            & $ 0.9090 \pm 0.06351$                   & 15.07G
       & 142.3K \\
NasN\_Brain                                                                  & $4.495 \pm 6.306$       & $2.032 \pm 2.579$        & $33.27 \pm 6.475$            & $ 0.9117 \pm 0.06286$                   & 14.39G
       & 135.6K \\
\hline
\end{tabular}
\end{table*}

\begin{figure}[!h]
\centerline{\includegraphics[width=\columnwidth]{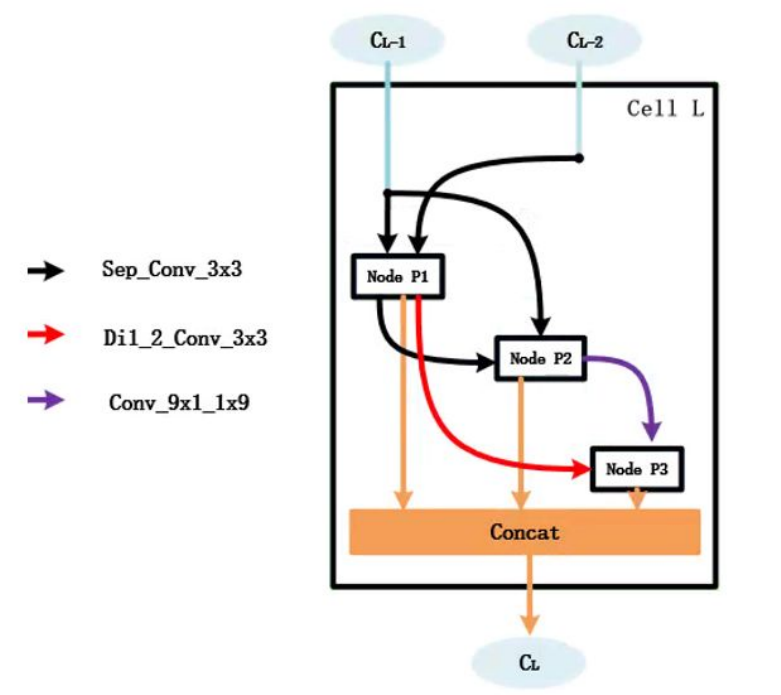}}
\caption{ The searched cell structure on the brain MR dataset, named NasN\_Brain.}
\label{fig16}
\end{figure}

We evaluated the generalizability of the searched architecture on the mini-fastMRI brain dataset from the following two aspects: for one, the architecture searched from the original knee data-set, i.e. NasN\_Knee, was directly generalized to the brain dataset via re-training; for the other, we searched for another specific architecture with the similar setting of NasN\_Knee for brain MR reconstruction, named NasN\_Brain, but with an extended searching space including all the six operation types mentioned in this manuscript. The searched architecture is shown in Fig.\ref{fig16}.

\begin{figure*}
\centerline{\includegraphics[width=18cm]{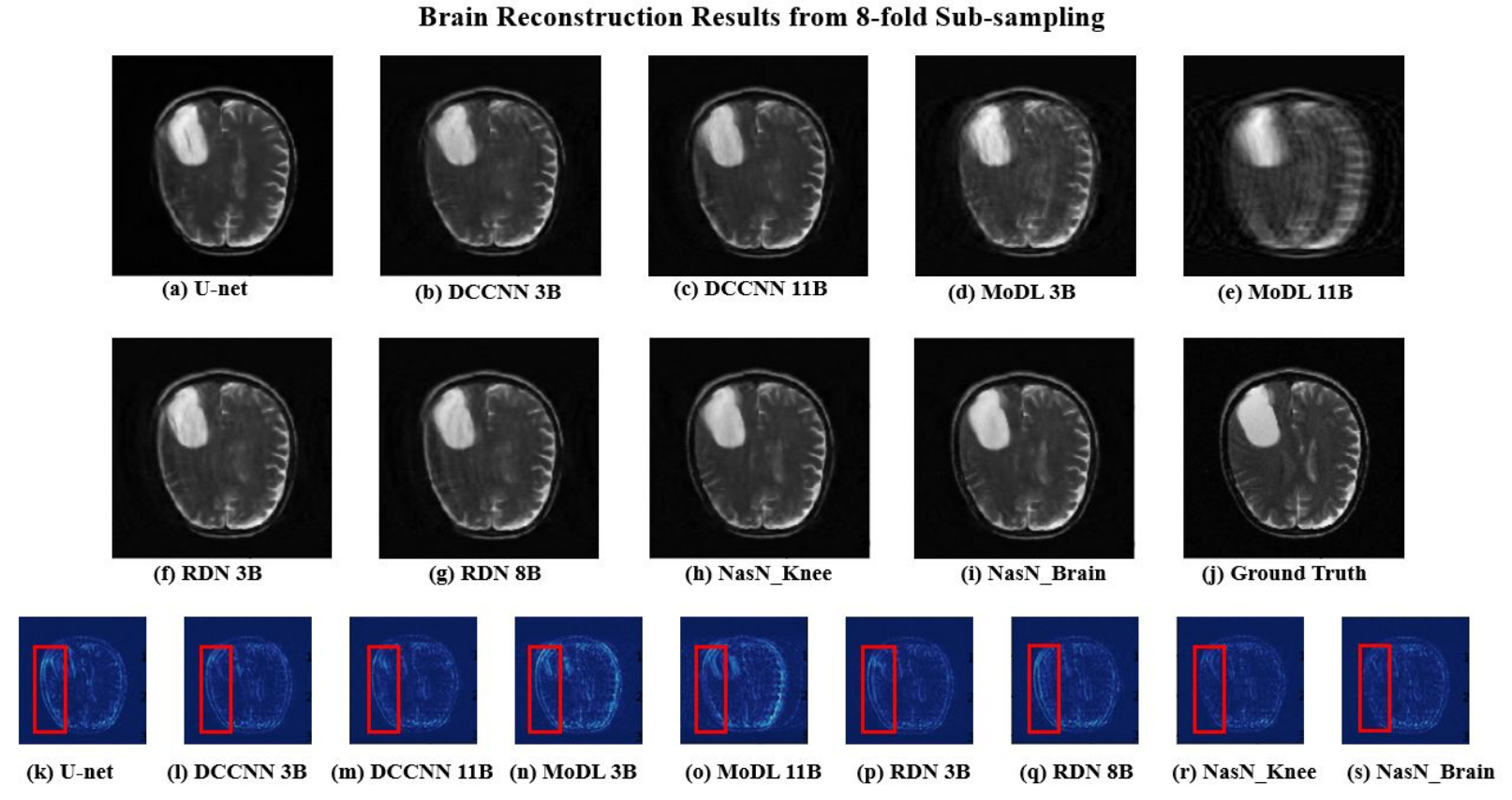}}
\caption{The qualitative brain MR reconstruction results of different DL based methods with 8-fold sub-sampling. Here the red boxes address where our reconstruction results have less noise.}
\label{fig17}
\end{figure*}

The quantitative brain MR reconstruction results are sho-wn in Tab.\ref{tab4}. We can find that NasN\_Knee re-trained on the brain dataset still exceeds previous baseline models, demonstrating that the searched architectures can be generalized to different tasks via re-training. And NasN\_Brain can outperform NasN\_Knee with fewer FLOPs and parameters, indicating that the architecture has a latent impact for MR reconstruction. The performance degradation 
of RDN with more blocks and MoDL implies that reducing learning parameters by sharing weights does not always work in different tasks. The qualitative brain MR reconstruction results of all DL based methods are shown in Fig.\ref{fig17} with 8-fold sub-sampling ratios.

We also re-trained NasN\_Brain on the knee dataset and it produces slightly worse results than NasN\_Knee but still outperforms manual designed baseline deep models shown in Tab.\ref{tab5}. We can conclude that the searched architecture is task-specific for MR reconstruction of different organs.

\begin{table*}[!t]
\caption{\label{tab5}Quantitative reconstruction performance of the searched architectures on different datasets. ($AVG \pm STD$)}
\centering
\begin{tabular}{p{2cm}|p{2cm}|p{2cm}|p{2cm}|p{2cm}|p{2.5cm}}
\hline
Dataset                & Model      & MSE ($\times$1e-10)       & NMSE($\times$1e-2)     & PSNR(dB)    & SSIM     \\
\hline
\multirow{2}{*}{Knee}  & NasN\_Knee  & $1.432 \pm 2.919$       & $5.112 \pm 6.408$        & $29.83 \pm 6.692$            & $ 0.6204 \pm 0.2676$                         \\
                       & NasN\_Brain & $1.454 \pm 2.996$       & $5.153 \pm 6.433$        & $29.78 \pm 6.696$            & $ 0.6202 \pm 0.2685$      \\
\hline
\multirow{2}{*}{Brain} & NasN\_Knee  & $4.596 \pm 6.236$       & $2.082 \pm 2.561$        & $33.12 \pm 6.348$            & $ 0.9090 \pm 0.06351$      \\
                       & NasN\_Brain & $4.495 \pm 6.306$       & $2.032 \pm 2.579$        & $33.27 \pm 6.475$            & $ 0.9117 \pm 0.06286$       \\
\hline
\end{tabular}
\end{table*}

\section{Discussion and Conclusion}

The MRI acceleration is highly demanded in clinical practice and has been an active research area for years. The introduction of compressed sensing made a significant breakthrough in the reduction of MRI scan time. Nowadays, deep learning technology brings new chances for us to solve this problem better. Although plenty of works have been explored, there still exists a gap between research works and clinical practice due to complex network architecture design and heavy computation cost. In this work, we evaluated the applicability of NAS to improve the DL based CS-MRI performance remarkably. The key insight of our work is that we can search for a specific and novel network architecture for CS-MRI in a differentiable way, based on continuous relaxation of the cell. Experiments demonstrate that our optimized network can reconstruct MR images better than previous handcrafted networks even with 6 times fewer computation resources. These results imply that automatically searched neural architecture can balance performance and efficiency better, so may be more friendly for possible clinical translation in the future.

The analysis of how hyper-parameters may affect the network architectures found by NAS shows that the searched structures follow some basic principles. Deeper feature maps in the single block require larger perception fields and should be properly refined with skip connection from the shallower layers. These principles may offer insights for networks used in other medical image tasks. And the setting of layer types in operation searching space makes it an open workflow allowing us to add other advanced structures in the future. Extensive experiments to explore the generalizability of the searched architectures prove that the searched network can be directly generalized to different organ MR reconstruction tasks via re-training. And if we re-search the architecture  for the exact application, the network will be more task-specific.

The following problems of this work will be explored in the near future. Due to the limited computation resources we have, we performed all experiments on single-coil MR reconstruction cases. But our methods can be generalized to multi-coil reconstruction by using 3D convolutional layers instead of 2D ones in the operation search space with the conjugate gradient solution instead of the close-formed one as MoDL\cite{aggarwal2018modl} mentioned in Sec.\ref{sec:form}. 
Another concern lies in the simple loss function, i.e. L1 loss and 1D Cartesian sub-sampling pattern adopted in this manuscript for demonstrating purpose, other loss functions, i.e. GAN\cite{mardani2018gan} loss, and sub-sampling patterns will be further explored with our workflow to improve the quality of final reconstruction.

To conclude, we present a novel reconstruction network for the CS-MRI task by automatically searching the inner structure. Experiments show the searched network can reconstruct MR images better and more efficiently than previous works. The searched architectures can be directly generalized to different MR reconstruction tasks of different organs through re-training and will be more task-specific via the architecture re-searching. With the superiority of good performance and the general applicability of neural architecture search, we expect that the proposed
workflow can become a promising research direction for MRI acceleration with great potential impacts on other medical image applications. 

\appendix

\printcredits

\section*{Funding}
This research was partly supported by National Natural Science Foundation of China (Grant No. 41876098 and Grant No. 81901734) and Overseas Cooperation Research Fund of Tsinghua Shenzhen International Graduate School. (Grant No. HW201808). 

\section*{Acknowledgment}
The authors would like to thank Dr. Zechen Zhou, Philips Research North America, Cambridge, Massachusetts, and Dr. Rui Li, Center for Biomedical Imaging Research, Department of Biomedical Engineering, School of Medicine, Tsinghua University for their valuable discussions that significantly improved the quality of the manuscript. They also thank the authors of fastMRI \cite{fastMRI}  for making their code and dataset accessible online. Special thanks are also given to the reviewers for their constructive comments.

%% Loading bibliography style file
%\bibliographystyle{model1-num-names}
\bibliographystyle{cas-model2-names}

% Loading bibliography database
\bibliography{cas-refs}

\begin{thebibliography}{61}
\expandafter\ifx\csname natexlab\endcsname\relax\def\natexlab#1{#1}\fi
\providecommand{\url}[1]{\texttt{#1}}
\providecommand{\href}[2]{#2}
\providecommand{\path}[1]{#1}
\providecommand{\DOIprefix}{doi:}
\providecommand{\ArXivprefix}{arXiv:}
\providecommand{\URLprefix}{URL: }
\providecommand{\Pubmedprefix}{pmid:}
\providecommand{\doi}[1]{\href{http://dx.doi.org/#1}{\path{#1}}}
\providecommand{\Pubmed}[1]{\href{pmid:#1}{\path{#1}}}
\providecommand{\bibinfo}[2]{#2}
\ifx\xfnm\relax \def\xfnm[#1]{\unskip,\space#1}\fi
%Type = Article
\bibitem[{Aggarwal et~al.(2018)Aggarwal, Mani and Jacob}]{aggarwal2018modl}
\bibinfo{author}{Aggarwal, H.K.}, \bibinfo{author}{Mani, M.P.},
  \bibinfo{author}{Jacob, M.}, \bibinfo{year}{2018}.
\newblock \bibinfo{title}{Modl: Model-based deep learning architecture for
  inverse problems}.
\newblock \bibinfo{journal}{IEEE transactions on medical imaging}
  \bibinfo{volume}{38}, \bibinfo{pages}{394--405}.
%Type = Article
\bibitem[{Ak{\c{c}}akaya et~al.(2019)Ak{\c{c}}akaya, Moeller, Weing{\"a}rtner
  and U{\u{g}}urbil}]{akccakaya2019LAKI}
\bibinfo{author}{Ak{\c{c}}akaya, M.}, \bibinfo{author}{Moeller, S.},
  \bibinfo{author}{Weing{\"a}rtner, S.}, \bibinfo{author}{U{\u{g}}urbil, K.},
  \bibinfo{year}{2019}.
\newblock \bibinfo{title}{Scan-specific robust artificial-neural-networks for
  k-space interpolation (raki) reconstruction: Database-free deep learning for
  fast imaging}.
\newblock \bibinfo{journal}{Magnetic resonance in medicine}
  \bibinfo{volume}{81}, \bibinfo{pages}{439--453}.
%Type = Article
\bibitem[{Angeline et~al.(1994)Angeline, Saunders and
  Pollack}]{angeline1994evolutionary}
\bibinfo{author}{Angeline, P.J.}, \bibinfo{author}{Saunders, G.M.},
  \bibinfo{author}{Pollack, J.B.}, \bibinfo{year}{1994}.
\newblock \bibinfo{title}{An evolutionary algorithm that constructs recurrent
  neural networks}.
\newblock \bibinfo{journal}{IEEE transactions on Neural Networks}
  \bibinfo{volume}{5}, \bibinfo{pages}{54--65}.
%Type = Article
\bibitem[{Baker et~al.(2017)Baker, Gupta, Raskar and
  Naik}]{baker2017accelerating}
\bibinfo{author}{Baker, B.}, \bibinfo{author}{Gupta, O.},
  \bibinfo{author}{Raskar, R.}, \bibinfo{author}{Naik, N.},
  \bibinfo{year}{2017}.
\newblock \bibinfo{title}{Accelerating neural architecture search using
  performance prediction}.
\newblock \bibinfo{journal}{arXiv preprint arXiv:1705.10823} .
%Type = Inproceedings
\bibitem[{Cai et~al.(2018)Cai, Chen, Zhang, Yu and Wang}]{cai2018efficient}
\bibinfo{author}{Cai, H.}, \bibinfo{author}{Chen, T.}, \bibinfo{author}{Zhang,
  W.}, \bibinfo{author}{Yu, Y.}, \bibinfo{author}{Wang, J.},
  \bibinfo{year}{2018}.
\newblock \bibinfo{title}{Efficient architecture search by network
  transformation}, in: \bibinfo{booktitle}{Thirty-Second AAAI Conference on
  Artificial Intelligence}.
%Type = Inproceedings
\bibitem[{Caruana et~al.(2000)Caruana, Lawrence and
  Giles}]{Caruana2000OverfittingIN}
\bibinfo{author}{Caruana, R.}, \bibinfo{author}{Lawrence, S.},
  \bibinfo{author}{Giles, C.L.}, \bibinfo{year}{2000}.
\newblock \bibinfo{title}{Overfitting in neural nets: Backpropagation,
  conjugate gradient, and early stopping}, in: \bibinfo{booktitle}{NIPS}.
%Type = Inproceedings
\bibitem[{Chollet(2017)}]{chollet2017xception}
\bibinfo{author}{Chollet, F.}, \bibinfo{year}{2017}.
\newblock \bibinfo{title}{Xception: Deep learning with depthwise separable
  convolutions}, in: \bibinfo{booktitle}{Proceedings of the IEEE conference on
  computer vision and pattern recognition}, pp. \bibinfo{pages}{1251--1258}.
%Type = Article
\bibitem[{Danelakis et~al.(2018)Danelakis, Theoharis and
  Verganelakis}]{danelakis2018survey}
\bibinfo{author}{Danelakis, A.}, \bibinfo{author}{Theoharis, T.},
  \bibinfo{author}{Verganelakis, D.A.}, \bibinfo{year}{2018}.
\newblock \bibinfo{title}{Survey of automated multiple sclerosis lesion
  segmentation techniques on magnetic resonance imaging}.
\newblock \bibinfo{journal}{Computerized Medical Imaging and Graphics}
  \bibinfo{volume}{70}, \bibinfo{pages}{83--100}.
%Type = Article
\bibitem[{Dong et~al.(2014)Dong, Loy, He and Tang}]{Dong2014ImageSU}
\bibinfo{author}{Dong, C.}, \bibinfo{author}{Loy, C.C.}, \bibinfo{author}{He,
  K.}, \bibinfo{author}{Tang, X.}, \bibinfo{year}{2014}.
\newblock \bibinfo{title}{Image super-resolution using deep convolutional
  networks}.
\newblock \bibinfo{journal}{IEEE Transactions on Pattern Analysis and Machine
  Intelligence} \bibinfo{volume}{38}, \bibinfo{pages}{295--307}.
%Type = Article
\bibitem[{Donoho(2006)}]{Donoho2006Compressed}
\bibinfo{author}{Donoho, D.L.}, \bibinfo{year}{2006}.
\newblock \bibinfo{title}{Compressed sensing}.
\newblock \bibinfo{journal}{IEEE Transactions on Information Theory}
  \bibinfo{volume}{52}, \bibinfo{pages}{1289--1306}.
%Type = Article
\bibitem[{Fiszelew et~al.(2007)Fiszelew, Britos, Ochoa, Merlino, Fern{\'a}ndez
  and Garc{\'\i}a-Mart{\'\i}nez}]{fiszelew2007finding}
\bibinfo{author}{Fiszelew, A.}, \bibinfo{author}{Britos, P.},
  \bibinfo{author}{Ochoa, A.}, \bibinfo{author}{Merlino, H.},
  \bibinfo{author}{Fern{\'a}ndez, E.},
  \bibinfo{author}{Garc{\'\i}a-Mart{\'\i}nez, R.}, \bibinfo{year}{2007}.
\newblock \bibinfo{title}{Finding optimal neural network architecture using
  genetic algorithms}.
\newblock \bibinfo{journal}{Advances in computer science and engineering
  research in computing science} \bibinfo{volume}{27}, \bibinfo{pages}{15--24}.
%Type = Inproceedings
\bibitem[{Girshick et~al.(2014)Girshick, Donahue, Darrell and
  Malik}]{Girshick2014Rich}
\bibinfo{author}{Girshick, R.}, \bibinfo{author}{Donahue, J.},
  \bibinfo{author}{Darrell, T.}, \bibinfo{author}{Malik, J.},
  \bibinfo{year}{2014}.
\newblock \bibinfo{title}{Rich feature hierarchies for accurate object
  detection and semantic segmentation}, in: \bibinfo{booktitle}{IEEE Conference
  on Computer Vision and Pattern Recognition}.
%Type = Inproceedings
\bibitem[{Goodfellow et~al.(2014)Goodfellow, Pouget-Abadie, Mirza, Bing,
  Warde-Farley, Ozair, Courville and Bengio}]{Goodfellow2014Generative}
\bibinfo{author}{Goodfellow, I.J.}, \bibinfo{author}{Pouget-Abadie, J.},
  \bibinfo{author}{Mirza, M.}, \bibinfo{author}{Bing, X.},
  \bibinfo{author}{Warde-Farley, D.}, \bibinfo{author}{Ozair, S.},
  \bibinfo{author}{Courville, A.}, \bibinfo{author}{Bengio, Y.},
  \bibinfo{year}{2014}.
\newblock \bibinfo{title}{Generative adversarial nets}, in:
  \bibinfo{booktitle}{International Conference on Neural Information Processing
  Systems}.
%Type = Article
\bibitem[{Hammernik et~al.(2018)Hammernik, Klatzer, Kobler, Recht, Sodickson,
  Pock and Knoll}]{hammernik2018vn}
\bibinfo{author}{Hammernik, K.}, \bibinfo{author}{Klatzer, T.},
  \bibinfo{author}{Kobler, E.}, \bibinfo{author}{Recht, M.P.},
  \bibinfo{author}{Sodickson, D.K.}, \bibinfo{author}{Pock, T.},
  \bibinfo{author}{Knoll, F.}, \bibinfo{year}{2018}.
\newblock \bibinfo{title}{Learning a variational network for reconstruction of
  accelerated mri data}.
\newblock \bibinfo{journal}{Magnetic resonance in medicine}
  \bibinfo{volume}{79}, \bibinfo{pages}{3055--3071}.
%Type = Inproceedings
\bibitem[{He et~al.(2016)He, Zhang, Ren and Jian}]{He2016Deep}
\bibinfo{author}{He, K.}, \bibinfo{author}{Zhang, X.}, \bibinfo{author}{Ren,
  S.}, \bibinfo{author}{Jian, S.}, \bibinfo{year}{2016}.
\newblock \bibinfo{title}{Deep residual learning for image recognition}, in:
  \bibinfo{booktitle}{IEEE Conference on Computer Vision and Pattern
  Recognition}.
%Type = Article
\bibitem[{Howard et~al.(2017)Howard, Zhu, Chen, Kalenichenko, Wang, Weyand,
  Andreetto and Adam}]{howard2017mobilenets}
\bibinfo{author}{Howard, A.G.}, \bibinfo{author}{Zhu, M.},
  \bibinfo{author}{Chen, B.}, \bibinfo{author}{Kalenichenko, D.},
  \bibinfo{author}{Wang, W.}, \bibinfo{author}{Weyand, T.},
  \bibinfo{author}{Andreetto, M.}, \bibinfo{author}{Adam, H.},
  \bibinfo{year}{2017}.
\newblock \bibinfo{title}{Mobilenets: Efficient convolutional neural networks
  for mobile vision applications}.
\newblock \bibinfo{journal}{arXiv preprint arXiv:1704.04861} .
%Type = Article
\bibitem[{Huang et~al.(2018)Huang, Yang, Wu, Qu, Yi and
  Metaxas}]{Huang2018MRIRV}
\bibinfo{author}{Huang, Q.}, \bibinfo{author}{Yang, D.}, \bibinfo{author}{Wu,
  P.}, \bibinfo{author}{Qu, H.}, \bibinfo{author}{Yi, J.},
  \bibinfo{author}{Metaxas, D.N.}, \bibinfo{year}{2018}.
\newblock \bibinfo{title}{Mri reconstruction via cascaded channel-wise
  attention network}.
\newblock \bibinfo{journal}{16th IEEE 16th International Symposium on
  Biomedical Imaging (ISBI 2019)} , \bibinfo{pages}{1622--1626}.
%Type = Article
\bibitem[{Ioffe and Szegedy(2015)}]{Ioffe2015BatchNA}
\bibinfo{author}{Ioffe, S.}, \bibinfo{author}{Szegedy, C.},
  \bibinfo{year}{2015}.
\newblock \bibinfo{title}{Batch normalization: Accelerating deep network
  training by reducing internal covariate shift}.
\newblock \bibinfo{journal}{ArXiv} \bibinfo{volume}{abs/1502.03167}.
%Type = Article
\bibitem[{Kim et~al.(2015)Kim, Lee and Lee}]{Kim2015DeeplyRecursiveCN}
\bibinfo{author}{Kim, J.}, \bibinfo{author}{Lee, J.K.}, \bibinfo{author}{Lee,
  K.M.}, \bibinfo{year}{2015}.
\newblock \bibinfo{title}{Deeply-recursive convolutional network for image
  super-resolution}.
\newblock \bibinfo{journal}{2016 IEEE Conference on Computer Vision and Pattern
  Recognition (CVPR)} , \bibinfo{pages}{1637--1645}.
%Type = Article
\bibitem[{Kim et~al.(2019)Kim, Garg and Haldar}]{kim2019loraki}
\bibinfo{author}{Kim, T.H.}, \bibinfo{author}{Garg, P.},
  \bibinfo{author}{Haldar, J.P.}, \bibinfo{year}{2019}.
\newblock \bibinfo{title}{Loraki: Autocalibrated recurrent neural networks for
  autoregressive mri reconstruction in k-space}.
\newblock \bibinfo{journal}{arXiv preprint arXiv:1904.09390} .
%Type = Article
\bibitem[{Kingma and Ba(2014)}]{kingma2014adam}
\bibinfo{author}{Kingma, D.P.}, \bibinfo{author}{Ba, J.}, \bibinfo{year}{2014}.
\newblock \bibinfo{title}{Adam: A method for stochastic optimization}.
\newblock \bibinfo{journal}{arXiv preprint arXiv:1412.6980} .
%Type = Article
\bibitem[{Krizhevsky and Hinton(2010)}]{krizhevsky2010convolutional}
\bibinfo{author}{Krizhevsky, A.}, \bibinfo{author}{Hinton, G.},
  \bibinfo{year}{2010}.
\newblock \bibinfo{title}{Convolutional deep belief networks on cifar-10}.
\newblock \bibinfo{journal}{Unpublished manuscript} \bibinfo{volume}{40},
  \bibinfo{pages}{1--9}.
%Type = Article
\bibitem[{LeCun et~al.(2015)LeCun, Bengio and Hinton}]{Lecun2015Deep}
\bibinfo{author}{LeCun, Y.}, \bibinfo{author}{Bengio, Y.},
  \bibinfo{author}{Hinton, G.}, \bibinfo{year}{2015}.
\newblock \bibinfo{title}{Deep learning}.
\newblock \bibinfo{journal}{nature} \bibinfo{volume}{521},
  \bibinfo{pages}{436--444}.
%Type = Inproceedings
\bibitem[{LeCun et~al.(1989)LeCun, Boser, Denker, Henderson, Howard, Hubbard
  and Jackel}]{LeCun1989HandwrittenDR}
\bibinfo{author}{LeCun, Y.}, \bibinfo{author}{Boser, B.E.},
  \bibinfo{author}{Denker, J.S.}, \bibinfo{author}{Henderson, D.},
  \bibinfo{author}{Howard, R.E.}, \bibinfo{author}{Hubbard, W.E.},
  \bibinfo{author}{Jackel, L.D.}, \bibinfo{year}{1989}.
\newblock \bibinfo{title}{Handwritten digit recognition with a back-propagation
  network}, in: \bibinfo{booktitle}{NIPS}.
%Type = Article
\bibitem[{Lee et~al.(2018)Lee, Yoo, Tak and Ye}]{lee2018deep}
\bibinfo{author}{Lee, D.}, \bibinfo{author}{Yoo, J.}, \bibinfo{author}{Tak,
  S.}, \bibinfo{author}{Ye, J.C.}, \bibinfo{year}{2018}.
\newblock \bibinfo{title}{Deep residual learning for accelerated mri using
  magnitude and phase networks}.
\newblock \bibinfo{journal}{IEEE Transactions on Biomedical Engineering}
  \bibinfo{volume}{65}, \bibinfo{pages}{1985--1995}.
%Type = Book
\bibitem[{Liang and Lauterbur(2000)}]{liang2000principles}
\bibinfo{author}{Liang, Z.P.}, \bibinfo{author}{Lauterbur, P.C.},
  \bibinfo{year}{2000}.
\newblock \bibinfo{title}{Principles of magnetic resonance imaging: a signal
  processing perspective}.
\newblock \bibinfo{publisher}{SPIE Optical Engineering Press}.
%Type = Inproceedings
\bibitem[{Liu et~al.(2019)Liu, Chen, Schroff, Adam, Hua, Yuille and
  Fei-Fei}]{liu2019auto}
\bibinfo{author}{Liu, C.}, \bibinfo{author}{Chen, L.C.},
  \bibinfo{author}{Schroff, F.}, \bibinfo{author}{Adam, H.},
  \bibinfo{author}{Hua, W.}, \bibinfo{author}{Yuille, A.L.},
  \bibinfo{author}{Fei-Fei, L.}, \bibinfo{year}{2019}.
\newblock \bibinfo{title}{Auto-deeplab: Hierarchical neural architecture search
  for semantic image segmentation}, in: \bibinfo{booktitle}{Proceedings of the
  IEEE Conference on Computer Vision and Pattern Recognition}, pp.
  \bibinfo{pages}{82--92}.
%Type = Inproceedings
\bibitem[{Liu et~al.(2018a)Liu, Zoph, Neumann, Shlens, Hua, Li, Fei-Fei,
  Yuille, Huang and Murphy}]{liu2018progressive}
\bibinfo{author}{Liu, C.}, \bibinfo{author}{Zoph, B.},
  \bibinfo{author}{Neumann, M.}, \bibinfo{author}{Shlens, J.},
  \bibinfo{author}{Hua, W.}, \bibinfo{author}{Li, L.J.},
  \bibinfo{author}{Fei-Fei, L.}, \bibinfo{author}{Yuille, A.},
  \bibinfo{author}{Huang, J.}, \bibinfo{author}{Murphy, K.},
  \bibinfo{year}{2018}a.
\newblock \bibinfo{title}{Progressive neural architecture search}, in:
  \bibinfo{booktitle}{Proceedings of the European Conference on Computer Vision
  (ECCV)}, pp. \bibinfo{pages}{19--34}.
%Type = Article
\bibitem[{Liu et~al.(2018b)Liu, Simonyan and Yang}]{liu2018darts}
\bibinfo{author}{Liu, H.}, \bibinfo{author}{Simonyan, K.},
  \bibinfo{author}{Yang, Y.}, \bibinfo{year}{2018}b.
\newblock \bibinfo{title}{Darts: Differentiable architecture search}.
\newblock \bibinfo{journal}{arXiv preprint arXiv:1806.09055} .
%Type = Article
\bibitem[{Liu et~al.(2017)Liu, Wang and Liang}]{liu2017sparse}
\bibinfo{author}{Liu, Q.}, \bibinfo{author}{Wang, S.}, \bibinfo{author}{Liang,
  D.}, \bibinfo{year}{2017}.
\newblock \bibinfo{title}{Sparse and dense hybrid representation via subspace
  modeling for dynamic mri}.
\newblock \bibinfo{journal}{Computerized Medical Imaging and Graphics}
  \bibinfo{volume}{56}, \bibinfo{pages}{24--37}.
%Type = Inproceedings
\bibitem[{Long et~al.(2015)Long, Shelhamer and Darrell}]{Long2015FullyCN}
\bibinfo{author}{Long, J.}, \bibinfo{author}{Shelhamer, E.},
  \bibinfo{author}{Darrell, T.}, \bibinfo{year}{2015}.
\newblock \bibinfo{title}{Fully convolutional networks for semantic
  segmentation}, in: \bibinfo{booktitle}{CVPR}.
%Type = Article
\bibitem[{Lustig et~al.(2008)Lustig, Donoho, Santos and
  Pauly}]{lustig2008compressed}
\bibinfo{author}{Lustig, M.}, \bibinfo{author}{Donoho, D.L.},
  \bibinfo{author}{Santos, J.M.}, \bibinfo{author}{Pauly, J.M.},
  \bibinfo{year}{2008}.
\newblock \bibinfo{title}{Compressed sensing mri}.
\newblock \bibinfo{journal}{IEEE signal processing magazine}
  \bibinfo{volume}{25}, \bibinfo{pages}{72}.
%Type = Article
\bibitem[{Mardani et~al.(2018)Mardani, Gong, Cheng, Vasanawala, Zaharchuk, Xing
  and Pauly}]{mardani2018gan}
\bibinfo{author}{Mardani, M.}, \bibinfo{author}{Gong, E.},
  \bibinfo{author}{Cheng, J.Y.}, \bibinfo{author}{Vasanawala, S.S.},
  \bibinfo{author}{Zaharchuk, G.}, \bibinfo{author}{Xing, L.},
  \bibinfo{author}{Pauly, J.M.}, \bibinfo{year}{2018}.
\newblock \bibinfo{title}{Deep generative adversarial neural networks for
  compressive sensing mri}.
\newblock \bibinfo{journal}{IEEE transactions on medical imaging}
  \bibinfo{volume}{38}, \bibinfo{pages}{167--179}.
%Type = Inproceedings
\bibitem[{Mikolov et~al.(2010)Mikolov, Karafi{\'a}t, Burget, {\v{C}}ernock{\`y}
  and Khudanpur}]{mikolov2010recurrent}
\bibinfo{author}{Mikolov, T.}, \bibinfo{author}{Karafi{\'a}t, M.},
  \bibinfo{author}{Burget, L.}, \bibinfo{author}{{\v{C}}ernock{\`y}, J.},
  \bibinfo{author}{Khudanpur, S.}, \bibinfo{year}{2010}.
\newblock \bibinfo{title}{Recurrent neural network based language model}, in:
  \bibinfo{booktitle}{Eleventh annual conference of the international speech
  communication association}.
%Type = Inproceedings
\bibitem[{Paszke et~al.(2019)Paszke, Gross, Massa, Lerer, Bradbury, Chanan,
  Killeen, Lin, Gimelshein, Antiga et~al.}]{paszke2019pytorch}
\bibinfo{author}{Paszke, A.}, \bibinfo{author}{Gross, S.},
  \bibinfo{author}{Massa, F.}, \bibinfo{author}{Lerer, A.},
  \bibinfo{author}{Bradbury, J.}, \bibinfo{author}{Chanan, G.},
  \bibinfo{author}{Killeen, T.}, \bibinfo{author}{Lin, Z.},
  \bibinfo{author}{Gimelshein, N.}, \bibinfo{author}{Antiga, L.}, et~al.,
  \bibinfo{year}{2019}.
\newblock \bibinfo{title}{Pytorch: An imperative style, high-performance deep
  learning library}, in: \bibinfo{booktitle}{Advances in Neural Information
  Processing Systems}, pp. \bibinfo{pages}{8024--8035}.
%Type = Inproceedings
\bibitem[{Peng et~al.(2017)Peng, Zhang, Yu, Luo and Sun}]{peng2017large}
\bibinfo{author}{Peng, C.}, \bibinfo{author}{Zhang, X.}, \bibinfo{author}{Yu,
  G.}, \bibinfo{author}{Luo, G.}, \bibinfo{author}{Sun, J.},
  \bibinfo{year}{2017}.
\newblock \bibinfo{title}{Large kernel matters--improve semantic segmentation
  by global convolutional network}, in: \bibinfo{booktitle}{Proceedings of the
  IEEE conference on computer vision and pattern recognition}, pp.
  \bibinfo{pages}{4353--4361}.
%Type = Article
\bibitem[{Pham et~al.(2018)Pham, Guan, Zoph, Le and Dean}]{pham2018efficient}
\bibinfo{author}{Pham, H.}, \bibinfo{author}{Guan, M.Y.},
  \bibinfo{author}{Zoph, B.}, \bibinfo{author}{Le, Q.V.},
  \bibinfo{author}{Dean, J.}, \bibinfo{year}{2018}.
\newblock \bibinfo{title}{Efficient neural architecture search via parameter
  sharing}.
\newblock \bibinfo{journal}{arXiv preprint arXiv:1802.03268} .
%Type = Article
\bibitem[{Quan et~al.(2018)Quan, Nguyen-Duc and Jeong}]{Quan2018CyclicLoss}
\bibinfo{author}{Quan, T.M.}, \bibinfo{author}{Nguyen-Duc, T.},
  \bibinfo{author}{Jeong, W.K.}, \bibinfo{year}{2018}.
\newblock \bibinfo{title}{Compressed sensing mri reconstruction using a
  generative adversarial network with a cyclic loss}.
\newblock \bibinfo{journal}{IEEE Transactions on Medical Imaging}
  \bibinfo{volume}{37}, \bibinfo{pages}{1488--1497}.
%Type = Inproceedings
\bibitem[{Real et~al.(2019)Real, Aggarwal, Huang and Le}]{real2019regularized}
\bibinfo{author}{Real, E.}, \bibinfo{author}{Aggarwal, A.},
  \bibinfo{author}{Huang, Y.}, \bibinfo{author}{Le, Q.V.},
  \bibinfo{year}{2019}.
\newblock \bibinfo{title}{Regularized evolution for image classifier
  architecture search}, in: \bibinfo{booktitle}{Proceedings of the AAAI
  Conference on Artificial Intelligence}, pp. \bibinfo{pages}{4780--4789}.
%Type = Inproceedings
\bibitem[{Real et~al.(2017)Real, Moore, Selle, Saxena, Suematsu, Tan, Le and
  Kurakin}]{real2017large}
\bibinfo{author}{Real, E.}, \bibinfo{author}{Moore, S.},
  \bibinfo{author}{Selle, A.}, \bibinfo{author}{Saxena, S.},
  \bibinfo{author}{Suematsu, Y.L.}, \bibinfo{author}{Tan, J.},
  \bibinfo{author}{Le, Q.V.}, \bibinfo{author}{Kurakin, A.},
  \bibinfo{year}{2017}.
\newblock \bibinfo{title}{Large-scale evolution of image classifiers}, in:
  \bibinfo{booktitle}{Proceedings of the 34th International Conference on
  Machine Learning-Volume 70}, \bibinfo{organization}{JMLR. org}. pp.
  \bibinfo{pages}{2902--2911}.
%Type = Inproceedings
\bibitem[{Ronneberger et~al.(2015)Ronneberger, Fischer and
  Brox}]{Ronneberger2015U}
\bibinfo{author}{Ronneberger, O.}, \bibinfo{author}{Fischer, P.},
  \bibinfo{author}{Brox, T.}, \bibinfo{year}{2015}.
\newblock \bibinfo{title}{U-net: Convolutional networks for biomedical image
  segmentation}, in: \bibinfo{booktitle}{International Conference on Medical
  Image Computing and Computer Assisted Intervention}.
%Type = Article
\bibitem[{Rudin et~al.(1992)Rudin, Osher and Fatemi}]{Rudin1992NonlinearTV}
\bibinfo{author}{Rudin, L.I.}, \bibinfo{author}{Osher, S.},
  \bibinfo{author}{Fatemi, E.}, \bibinfo{year}{1992}.
\newblock \bibinfo{title}{Nonlinear total variation based noise removal
  algorithms}.
\newblock \bibinfo{journal}{Physica D: nonlinear phenomena}
  \bibinfo{volume}{60}, \bibinfo{pages}{259--268}.
%Type = Article
\bibitem[{Schlemper et~al.(2017)Schlemper, Caballero, Hajnal, Price and
  Rueckert}]{Schlemper2017ADC}
\bibinfo{author}{Schlemper, J.}, \bibinfo{author}{Caballero, J.},
  \bibinfo{author}{Hajnal, J.V.}, \bibinfo{author}{Price, A.N.},
  \bibinfo{author}{Rueckert, D.}, \bibinfo{year}{2017}.
\newblock \bibinfo{title}{A deep cascade of convolutional neural networks for
  dynamic mr image reconstruction}.
\newblock \bibinfo{journal}{IEEE Transactions on Medical Imaging}
  \bibinfo{volume}{37}, \bibinfo{pages}{491--503}.
%Type = Inproceedings
\bibitem[{Sun et~al.(2016)Sun, Li, Xu et~al.}]{sun2016ADMM}
\bibinfo{author}{Sun, J.}, \bibinfo{author}{Li, H.}, \bibinfo{author}{Xu, Z.},
  et~al., \bibinfo{year}{2016}.
\newblock \bibinfo{title}{Deep admm-net for compressive sensing mri}, in:
  \bibinfo{booktitle}{Advances in neural information processing systems}, pp.
  \bibinfo{pages}{10--18}.
%Type = Inproceedings
\bibitem[{Sun et~al.(2018)Sun, Fan, Huang, Ding and
  Paisley}]{Sun2018CompressedSM}
\bibinfo{author}{Sun, L.}, \bibinfo{author}{Fan, Z.}, \bibinfo{author}{Huang,
  Y.}, \bibinfo{author}{Ding, X.}, \bibinfo{author}{Paisley, J.W.},
  \bibinfo{year}{2018}.
\newblock \bibinfo{title}{Compressed sensing mri using a recursive dilated
  network}, in: \bibinfo{booktitle}{AAAI}.
%Type = Book
\bibitem[{Sutton et~al.(1998)Sutton, Barto et~al.}]{sutton1998introduction}
\bibinfo{author}{Sutton, R.S.}, \bibinfo{author}{Barto, A.G.}, et~al.,
  \bibinfo{year}{1998}.
\newblock \bibinfo{title}{Introduction to reinforcement learning}.
  volume~\bibinfo{volume}{2}.
\newblock \bibinfo{publisher}{MIT press Cambridge}.
%Type = Inproceedings
\bibitem[{Szegedy et~al.(2017)Szegedy, Ioffe, Vanhoucke and
  Alemi}]{szegedy2017inception}
\bibinfo{author}{Szegedy, C.}, \bibinfo{author}{Ioffe, S.},
  \bibinfo{author}{Vanhoucke, V.}, \bibinfo{author}{Alemi, A.A.},
  \bibinfo{year}{2017}.
\newblock \bibinfo{title}{Inception-v4, inception-resnet and the impact of
  residual connections on learning}, in: \bibinfo{booktitle}{Thirty-First AAAI
  Conference on Artificial Intelligence}.
%Type = Article
\bibitem[{Tsao and Kozerke(2012)}]{tsao2012mri}
\bibinfo{author}{Tsao, J.}, \bibinfo{author}{Kozerke, S.},
  \bibinfo{year}{2012}.
\newblock \bibinfo{title}{Mri temporal acceleration techniques}.
\newblock \bibinfo{journal}{Journal of Magnetic Resonance Imaging}
  \bibinfo{volume}{36}, \bibinfo{pages}{543--560}.
%Type = Inproceedings
\bibitem[{Uecker et~al.(2015)Uecker, Ong, Tamir, Bahri, Virtue, Cheng, Zhang
  and Lustig}]{BART}
\bibinfo{author}{Uecker, M.}, \bibinfo{author}{Ong, F.},
  \bibinfo{author}{Tamir, J.I.}, \bibinfo{author}{Bahri, D.},
  \bibinfo{author}{Virtue, P.}, \bibinfo{author}{Cheng, J.Y.},
  \bibinfo{author}{Zhang, T.}, \bibinfo{author}{Lustig, M.},
  \bibinfo{year}{2015}.
\newblock \bibinfo{title}{Berkeley advanced reconstruction toolbox}, in:
  \bibinfo{booktitle}{Proc. Intl. Soc. Mag. Reson. Med}.
%Type = Inproceedings
\bibitem[{Wang et~al.(2016)Wang, Su, Ying, Xi, Zhu, Feng, Feng and
  Dong}]{Wang2016Accelerating}
\bibinfo{author}{Wang, S.}, \bibinfo{author}{Su, Z.}, \bibinfo{author}{Ying,
  L.}, \bibinfo{author}{Xi, P.}, \bibinfo{author}{Zhu, S.},
  \bibinfo{author}{Feng, L.}, \bibinfo{author}{Feng, D.},
  \bibinfo{author}{Dong, L.}, \bibinfo{year}{2016}.
\newblock \bibinfo{title}{Accelerating magnetic resonance imaging via deep
  learning}, in: \bibinfo{booktitle}{IEEE International Symposium on Biomedical
  Imaging}.
%Type = Inproceedings
\bibitem[{Xie et~al.(2012)Xie, Xu and Chen}]{Xie2012Image}
\bibinfo{author}{Xie, J.}, \bibinfo{author}{Xu, L.}, \bibinfo{author}{Chen,
  E.}, \bibinfo{year}{2012}.
\newblock \bibinfo{title}{Image denoising and inpainting with deep neural
  networks}, in: \bibinfo{booktitle}{International Conference on Neural
  Information Processing Systems}.
%Type = Article
\bibitem[{Yang et~al.(2017)Yang, Yu, Dong, Slabaugh, Dragotti, Ye, Liu,
  Arridge, Keegan and Guo}]{Yang2017DAGAN}
\bibinfo{author}{Yang, G.}, \bibinfo{author}{Yu, S.}, \bibinfo{author}{Dong,
  H.}, \bibinfo{author}{Slabaugh, G.}, \bibinfo{author}{Dragotti, P.L.},
  \bibinfo{author}{Ye, X.}, \bibinfo{author}{Liu, F.},
  \bibinfo{author}{Arridge, S.}, \bibinfo{author}{Keegan, J.},
  \bibinfo{author}{Guo, Y.}, \bibinfo{year}{2017}.
\newblock \bibinfo{title}{Dagan: Deep de-aliasing generative adversarial
  networks for fast compressed sensing mri reconstruction}.
\newblock \bibinfo{journal}{IEEE Transactions on Medical Imaging}
  \bibinfo{volume}{37}, \bibinfo{pages}{1310--1321}.
%Type = Article
\bibitem[{Yu and Koltun(2015)}]{Yu2015MultiScaleCA}
\bibinfo{author}{Yu, F.}, \bibinfo{author}{Koltun, V.}, \bibinfo{year}{2015}.
\newblock \bibinfo{title}{Multi-scale context aggregation by dilated
  convolutions}.
\newblock \bibinfo{journal}{ArXiv} \bibinfo{volume}{abs/1511.07122}.
%Type = Article
\bibitem[{Zbontar et~al.(2018)Zbontar, Knoll, Sriram, Muckley, Bruno, Defazio,
  Parente, Geras, Katsnelson, Chandarana, Zhang, Drozdzal, Romero, Rabbat,
  Vincent, Pinkerton, Wang, Yakubova, Owens, Zitnick, Recht, Sodickson and
  Lui}]{fastMRI}
\bibinfo{author}{Zbontar, J.}, \bibinfo{author}{Knoll, F.},
  \bibinfo{author}{Sriram, A.}, \bibinfo{author}{Muckley, M.J.},
  \bibinfo{author}{Bruno, M.}, \bibinfo{author}{Defazio, A.},
  \bibinfo{author}{Parente, M.}, \bibinfo{author}{Geras, K.},
  \bibinfo{author}{Katsnelson, J.}, \bibinfo{author}{Chandarana, H.},
  \bibinfo{author}{Zhang, Z.}, \bibinfo{author}{Drozdzal, M.},
  \bibinfo{author}{Romero, A.}, \bibinfo{author}{Rabbat, M.G.},
  \bibinfo{author}{Vincent, P.}, \bibinfo{author}{Pinkerton, J.},
  \bibinfo{author}{Wang, D.}, \bibinfo{author}{Yakubova, N.},
  \bibinfo{author}{Owens, E.}, \bibinfo{author}{Zitnick, C.L.},
  \bibinfo{author}{Recht, M.P.}, \bibinfo{author}{Sodickson, D.K.},
  \bibinfo{author}{Lui, Y.W.}, \bibinfo{year}{2018}.
\newblock \bibinfo{title}{fastmri: An open dataset and benchmarks for
  accelerated mri}.
\newblock \bibinfo{journal}{ArXiv} \bibinfo{volume}{abs/1811.08839}.
%Type = Article
\bibitem[{Zeng et~al.(2019)Zeng, Yang, Xiao and Chen}]{Zeng2019AVD}
\bibinfo{author}{Zeng, K.}, \bibinfo{author}{Yang, Y.}, \bibinfo{author}{Xiao,
  G.}, \bibinfo{author}{Chen, Z.}, \bibinfo{year}{2019}.
\newblock \bibinfo{title}{A very deep densely connected network for compressed
  sensing mri}.
\newblock \bibinfo{journal}{IEEE Access} \bibinfo{volume}{7},
  \bibinfo{pages}{85430--85439}.
%Type = Inproceedings
\bibitem[{Zhang et~al.(2018)Zhang, Zhou, Lin and Sun}]{zhang2018shufflenet}
\bibinfo{author}{Zhang, X.}, \bibinfo{author}{Zhou, X.}, \bibinfo{author}{Lin,
  M.}, \bibinfo{author}{Sun, J.}, \bibinfo{year}{2018}.
\newblock \bibinfo{title}{Shufflenet: An extremely efficient convolutional
  neural network for mobile devices}, in: \bibinfo{booktitle}{Proceedings of
  the IEEE Conference on Computer Vision and Pattern Recognition}, pp.
  \bibinfo{pages}{6848--6856}.
%Type = Article
\bibitem[{Zhao et~al.(2017)Zhao, Gallo, Frosio and Kautz}]{Zhao2017Loss}
\bibinfo{author}{Zhao, H.}, \bibinfo{author}{Gallo, O.},
  \bibinfo{author}{Frosio, I.}, \bibinfo{author}{Kautz, J.},
  \bibinfo{year}{2017}.
\newblock \bibinfo{title}{Loss functions for image restoration with neural
  networks}.
\newblock \bibinfo{journal}{IEEE Transactions on Computational Imaging}
  \bibinfo{volume}{3}, \bibinfo{pages}{47--57}.
%Type = Article
\bibitem[{Zhao et~al.(2020)Zhao, Liu, Liu and Wong}]{zhao2020super}
\bibinfo{author}{Zhao, M.}, \bibinfo{author}{Liu, X.}, \bibinfo{author}{Liu,
  H.}, \bibinfo{author}{Wong, K.K.}, \bibinfo{year}{2020}.
\newblock \bibinfo{title}{Super-resolution of cardiac magnetic resonance images
  using laplacian pyramid based on generative adversarial networks}.
\newblock \bibinfo{journal}{Computerized Medical Imaging and Graphics}
  \bibinfo{volume}{80}, \bibinfo{pages}{101--698}.
%Type = Article
\bibitem[{Zhou et~al.(2004)Zhou, Alan~Conrad, Hamid~Rahim and
  Simoncelli}]{Zhou2004Image}
\bibinfo{author}{Zhou, W.}, \bibinfo{author}{Alan~Conrad, B.},
  \bibinfo{author}{Hamid~Rahim, S.}, \bibinfo{author}{Simoncelli, E.P.},
  \bibinfo{year}{2004}.
\newblock \bibinfo{title}{Image quality assessment: from error visibility to
  structural similarity}.
\newblock \bibinfo{journal}{IEEE Trans Image Process} \bibinfo{volume}{13},
  \bibinfo{pages}{600--612}.
%Type = Article
\bibitem[{Zhu et~al.(2018)Zhu, Liu, Cauley, Rosen and Rosen}]{zhu2018automap}
\bibinfo{author}{Zhu, B.}, \bibinfo{author}{Liu, J.Z.},
  \bibinfo{author}{Cauley, S.F.}, \bibinfo{author}{Rosen, B.R.},
  \bibinfo{author}{Rosen, M.S.}, \bibinfo{year}{2018}.
\newblock \bibinfo{title}{Image reconstruction by domain-transform manifold
  learning}.
\newblock \bibinfo{journal}{Nature} \bibinfo{volume}{555},
  \bibinfo{pages}{487}.
%Type = Article
\bibitem[{Zoph and Le(2016)}]{zoph2016neural}
\bibinfo{author}{Zoph, B.}, \bibinfo{author}{Le, Q.V.}, \bibinfo{year}{2016}.
\newblock \bibinfo{title}{Neural architecture search with reinforcement
  learning}.
\newblock \bibinfo{journal}{arXiv preprint arXiv:1611.01578} .

\end{thebibliography}

\end{document}